\documentclass[useAMS,usenatbib]{mn2e}
\usepackage{graphicx,epsfig,amssymb,amsmath,layout,verbatim,rotating,calc,
  mathrsfs,natbib}
\usepackage{graphicx,amssymb,amsmath,layout,verbatim,rotating,calc,mathrsfs,natbib}
\usepackage{graphics}
\usepackage{rotate}
\usepackage{float}
\usepackage{multirow}
\usepackage[dvipsnames]{xcolor}
\usepackage{soul}
\usepackage{color}
\usepackage{longtable}
\usepackage{arydshln}
\usepackage[normalem]{ulem}
\usepackage{txfonts}
\newcommand{\ej}{E$_{J}$ }

\title[Orbits in non-autonomous Hamiltonian
    systems]{Orbit evolution in growing stellar bars: Bar-supporting orbits at
the vertical ILR region}
\author[T.~Manos, Ch.~Skokos, P.A.~Patsis]
{T.~Manos,$^{1}$
Ch.~Skokos,$^{2}$
P.A.~Patsis,$^{3}$\thanks{patsis@academyofathens.gr}\\
$^1$Laboratoire de Physique Th{\'e}orique et Mod{\'e}lisation, CY
Cergy Paris Universit{\'e}, CNRS, UMR 8089, 95302 Cergy-Pontoise
cedex, France\\
$^2$Nonlinear Dynamics and Chaos group, Department of Mathematics and
Applied Mathematics, University of Cape Town, Rondebosch, 7701, Cape
Town, South Africa\\
$^3$Research Center for Astronomy, Academy of Athens, Soranou Efessiou
4, GR-115 27, Athens, Greece\\
}
\date{Accepted ..........Received .............;in original form ..........}

\begin{document}
\maketitle

\label{firstpage}

\begin{abstract}

We investigate the evolution of orbital shapes at the Inner Lindblad Resonance
region of a rotating three-dimensional bar, the mass of which is growing with
time. We evaluate in time-dependent models, during a 5~Gyr period, the
importance of orbits with initial conditions known to play a
significant role in supporting peanut-like structures in autonomous systems.
These orbits are the central family of periodic orbits (x1) and vertical
perturbations of it, orbits of its standard three-dimensional bifurcations at
the region (x1v1 and x1v2), as well as orbits in their neighbourhood. The
knowledge of the regular or chaotic character of these orbits is essential as
well, because it allows us to estimate their contribution to the support of a
rotating bar and, more importantly, the dynamical mechanisms that make it
possible. This is calculated by means of the GALI$_2$ index. We find that
orbital patterns existing in the autonomous case, persist for longer times in
the more massive bar models, and even more so in a model in which the
central spheroid component of our adopted galactic potential becomes rather
insignificant. The peanut-supporting orbits which we find, have a regular or, in
 most cases, a weakly chaotic character. There are cases in which orbits
starting  close to unstable periodic orbits in an autonomous model behave as
regular and support the bar when its mass increases with time. As a rule of
thumb for the orbital dynamics of our non-autonomous models at a certain time,
can be considered the dynamics of the corresponding frozen systems around that
time.
\end{abstract}


\section{Introduction}
\label{sec:intro}

\subsection{Background}
Over the years we have a fairly good understanding of the orbital dynamics in
rotating ellipsoids that model galactic bars. The potentials in this kind of
models do not change in time, thus in order to study them, the formalism of
autonomous Hamiltonian systems has been adopted  \citep[see e.g.][]{cp80, a83,
cg89, pf84, cm85, p05, spa02a, path19}.

In all the above studies, the main mechanism reinforcing the bar, is the
trapping of quasi-periodic orbits around stable periodic ones. In
two-dimensional systems, these stable periodic orbits are the elliptical-like
members of the x1 family \citep[see e.g.][]{cg89}, while in three-dimensional
models they are the stable periodic orbits of the x1-tree \citep{spa02a}.
According to the theory of dynamical systems, around a stable periodic orbit
there is a volume in phase space, where motion is regular, i.e. the orbits will
be quasi-periodic. In such a case, a particle following a quasi-periodic orbit,
will remain in the neighbourhood of the periodic orbit reinforcing in this way
the local density \citep[see e.g][sections 2.4 and 2.5]{gcobook}.

Nevertheless, not every quasi-periodic orbit in a model is bar-supporting. For
this, it has to enhance locally the density in such a way, as to reinforce the
morphological feature to be modeled, i.e. in our case the bar. A known
counterexample is the case of the nearly circular retrograde orbits of the
family x4, which remain stable for almost all Jaccobi constants in standard
barred galaxy models \citep[see e.g.][]{cp80}. Such orbits, if populated, would
lead to the appearance of sizable counter-rotating discs at the centers of the
bars, which are not observed. Most importantly there is the class of weakly
chaotic and sticky chaotic orbits \citep{ch08}, which during a certain time, may
also enhance a particular bar structure \citep{paq97}. As a result the
bar-supporting regions on a Poincar\'{e} surface of section, do not necessarily
correspond to those occupied by stability islands. The topologies  of the
regular and bar-supporting regions do not coincide \citep{cpb11}.

The next step is to examine what a time dependence of the potential may cause in
the orbital dynamics of barred galaxy models. A slow variation of the
gravitational field is encountered during the evolution of several $N$-body
models, which simulate galactic bars \citep[see e.g.][]{ath03,hk09}. ``Slow'',
means that the test particles have at least enough time to feel resonances
(radial and vertical), the location of which does not change considerably on the
galactic disks, over the time we consider. In the above mentioned work the
variation of the potential between snapshots with a time distance of a few~Gyr
has been found  to be small, so that the evolution of the model during this
period could reliably be approximated by a stationary mean gravitational field.

In \citet{mm14} and \citet{mm16}, the evolution of an $N$-body simulation of a
disc galaxy within a live halo \citep{ma10}, which results to the formation of a
strong bar, has been approximated by means of a time-dependent (TD) analytical
model. This TD model was composed of three components, namely a bar, a disc and
a halo. After the initial formation of a bar, there is a relative fast increase
of its size and strength, before the model enters a phase of slower variation of
the parameters of its components.

In the present work we focus in the vertical Inner Lindblad Resonance (vILR)
region of rotating bars. This region is either very close to, or practically
identical with, the radial Inner Lindblad Resonance (rILR) region in many
$N$-body, or analytic models \citep[e.g.][]{cbfp90, pk14a, pk14b}. We will refer
to it altogether as the ``ILR region''. In principle the ILR region combines
orbital content that could support the thin and the thick  part of the bar.

The goal of this study is to find out whether or not time variation of the
potential changes significantly the orbital content of the model. This would
mean that the observed structure of the  boxy-peanut (hereafter b/p) bulge
would be supported by totally different orbits than the known ones associated
with x1 and its three-dimensional (3D) bifurcations \citep{spa02a, spa02b,
psa02, pk14a, pk14b, ph18}. In addition, we want to investigate whether specific
orbits that support the bar in a time-independent (TI) model remain
``bar-supporting'', when a parameter like the mass of the bar ($M_B$) varies.

\subsection{Quantifying the chaoticity of the orbits}
\label{sec:qchaos}

Since the reinforcement of specific structures in TI potentials is associated
either with order or, under certain conditions, with weak chaos and stickiness
\citep{ch08}, it is important to know how regular or chaotic are the
bar-supporting orbits in our models. To estimate this, we use the GALI$_2$ index
 \citep{SBA07, SBA08, MSA12, MMS20}. The results have been compared with those
obtained for the same reason with the Maximal Lyapunov Exponent (MLE)
\citep{benetal80a, benetal80b, S10} of the orbits.

The GALI$_2$ index is given by the norm of the wedge product of two normalized
to unity deviation vectors $\mathbf{\hat{w}}_{1}(t)$ and
$\mathbf{\hat{w}}_{2}(t)$ from the studied orbit,
i.e.~$\rm{GALI}_{2}(t)=\|\mathbf{\hat{w}}_{1}(t)\wedge
\mathbf{\hat{w}}_{2}(t)\|.$ The initial coordinates, of the deviations
vectors are chosen randomly, and the two vectors are orthonormalized at the
beginning of the integration, setting in this way the initial value of the index
to $\mbox{GALI}_2(0)=1$. Thus, in order to evaluate GALI$_2$ we simultaneously
integrate the equations of motion and the so-called variational equations
\citep[see e.g.][]{S10}, which govern the evolution of the two deviation
vectors.

Let us briefly recall the behavior of the GALI$_2$ index \citep[see][and
references therein]{sm16}. For chaotic orbits the index falls exponentially fast
to zero as $\mbox{GALI}_{2}(t)\propto \exp{(-(\lambda_{1}-\lambda_{2})t)}$,
where $\lambda_{1}$ and $\lambda_{2}$ are the two largest Lyapunov exponents
\citep[for the definitions and for the computation of the Lyapunov exponents see
e.g.][]{benetal80a, benetal80b, S10}. On the other hand, for regular orbits it
oscillates around a positive value across the integration,
i.e.~$\mbox{GALI}_{2}(t)\propto \mbox{constant}$. In the case of weakly chaotic
or sticky orbits we observe a transition from practically constant GALI$_2$
values, which correspond to the seemingly quasiperiodic epoch of the orbit, to
an exponential decay to zero, which indicates the orbit's transition to
chaoticity.

Orbits in TI Hamiltonian systems are either regular or chaotic, which means that
their $\mbox{GALI}_{2}$ values will, respectively, oscillate around a constant
positive number or eventually become zero. On the other hand orbits in TD
potentials can exhibit more complicated behaviours as transitions between
regular and chaotic epochs in their evolution can be observed, depending on
their location in the changing phase space of the system. We use
$\mbox{GALI}_{2}$ to capture these dynamical changes of orbits in TD
Hamiltonians by applying the following procedure: Whenever $\mbox{GALI}_{2}$
reaches a very small value (namely when $\mbox{GALI}_{2} \leq 10^{-8}$) we
reinitialize its computation by taking again two new random orthonormal
deviation vectors (i.e.~setting anew $\mbox{GALI}_2=1$) and then let these
vectors evolve under the current dynamics. Since an exponential decrease of
$\mbox{GALI}_{2}$ indicates chaotic behaviour, successive and frequent
reinitializations of the index identify chaotic epochs, while extended periods
of time where $\mbox{GALI}_{2} > 10^{-8}$,  correspond to a regular behaviour
\citep{mbs13, mm14}.

In order to achieve the goals of our study we proceed as follows: We consider a
time interval of 5~Gyr, within which we perform our calculations. This is about
half the age of a Milky-Way-type galaxy. For this period we integrate and
characterize the orbits according to their chaoticity indices, as the bar mass
increases linearly in time from M$_{Bmin}$ to M$_{Bmax}$. We do so with a number
of characteristic orbits, which we know in advance that play a key role in
supporting bars
in TI potentials, i.e.~orbits associated with the families of the x1-tree
\citep{spa02a}. We check if they continue to support the initial bar structure
as they evolve, if they support a similar structure with different dimensions,
if they destroy the bar, or if they are transformed to other orbital shapes. In
the latter case, we check also if the new orbital shapes can be identified with
known patterns encountered in the phase space of TI systems. A significant
factor that determines the morphological evolution of each orbit is the rate of
variation of the TD parameter. For this purpose, we study for each considered
case the orbital evolution in a fast and in a slowly evolving potential.

In Section~\ref{models}, we describe the models we use, in Section~\ref{mA} we
study the behaviour of orbits in a potential with a relatively low mass bar,
while in Section~\ref{mB} we present the results of the corresponding study in a
more massive bar model. In Section~\ref{mC}, we  investigate the behaviour of
orbits in a special case, in which the evolution of the orbital stability of the
main 3D families of periodic orbits (POs) at the ILR region do not have complex
unstable \citep[for a definition see e.g.][]{spa02a} parts. Finally in
Section~\ref{sec:concl} we present and discuss our conclusions.

\section{The set-up of the models}
\label{models}

A 3D autonomous Hamiltonian system describing the
dynamics of a disc galaxy with a rotating bar  can be written, in
Cartesian coordinates $(x,y,z)$, in the form:
\begin{equation}
H= \frac{1}{2}(p_{x}^{2} + p_{y}^{2} + p_{z}^{2}) +
    \Phi(x,y,z) - \Omega_{b}(x p_{y} - y p_{x})    ,
\label{eq:ham}
\end{equation}
where $p_{x},~ p_{y},$ and $p_{z}$ are the canonically conjugate
momenta, $\Phi$ the gravitational potential of the model and
$\Omega_{b}$ the angular velocity of the system, in our case the
pattern speed of the bar. The numerical value of the Hamiltonian,
E$_J$ (Jacobi constant), is constant and we will also refer to it
throughout the paper as the ``energy''.

For the sake of continuity with our previous studies on the subject,
we use again in this paper the popular triaxial Ferrers bar model
\citep{f887}, which is described in detail in \citet{spa02a} and
\citet{pk14a}, with parameters close to those in the pioneer paper by
\citet{pf84}. The formulae for the axisymmetric part of the potential,
as well as the bar model, can be found in these references.

The Ferrers bar is inhomogeneous, with index 2 and axial ratios $a:b:c =
6:1.5:0.6$ (with $a$, $b$, $c$ being the semi-axes). We have taken as major axis
the $y-$axis. For our orbital calculations, i.e. for finding the periodic orbits
and  calculating Poincar\'{e} surfaces of sections \citep{poin99}, we consider
upwards intersections with the y=0 plane. The axisymmetric background consists
of a Miyamoto disc \citep{mn75} with fixed horizontal and vertical scale lengths
A=3 and B=1 respectively and a Plummer sphere \citep{pl11} representing the
bulge, with scale length $\epsilon_s=0.4$. The length unit is taken as 1~kpc,
the time unit as 1~Myr and the mass
unit as $2\times 10^{11} M_{\odot}$.

The masses of the three components satisfy \(G(M_{D}+M_{S}+M_{B})=1 \), where \(
M_{D} \) is the total mass of the disc, \(M_{S} \) the mass of the bulge
(spheroid), \( M_{B} \) is the mass of the bar component and $G$ the
gravitational  constant. We note that in our simulations the bar mass
increases at the expense of the disk mass, i.e.~the term $GM{_D}$ decreases
appropriately so that the condition \(G(M_{D}+M_{S}+M_{B})=1 \) is always
satisfied. An explicit halo component is not included, since our study refers to
the inner parts of the galaxy, where it is considered to be not important.

We investigate the orbital evolution in the following three models, aiming to
cover some typical and representative cases:
\begin{enumerate}
\item Model A: A bar model, in which $GM_{B}=0.05$, i.e. a relatively low mass
bar, and
  $GM_{S}=0.08$. In this model, the variation with \ej of the
  stability of the main simple periodic families encountered in 3D
  rotating bars, i.e. x1, x1v1 and x1v2, is typical for this kind of
  systems \citep{spa02a}.
\item Model B: A more massive bar model, with
  $GM_{B}=0.13$ and $GM_{S}=0.08$, in which we have again the usual
  variation of the stability indices of the three families.
\item Model C: A model in which the x1v1 family has no complex unstable parts
(see section~\ref{mA_starpo} below), all the way to corotation. In this case
$GM_{B}=0.1$ and $GM_{S}=0.022$.
\end{enumerate}

For all three models we follow the same steps, namely, we first investigate the
behaviour of the model in the autonomous case. Then, based on the topology of
the phase space at properly chosen energies, we follow the evolution of selected
orbits as the mass of the bar varies. These orbits have been found to play a
major role in supporting the b/p component in TI, rotating barred potentials.
The criterion for choosing the initial energy of the orbits in each model, is to
be in the ILR region, just beyond the critical \ej value for which the x1, x1v1
and x1v2 families are already present in the system. This procedure will be
described in more detail in model A and will be repeated for models B and C.

\section{Model A: A low mass bar}
\label{mA}
\subsection{The autonomous case}
\label{mA_autoc}

Our goal is to investigate the orbital evolution in galactic bars as the mass of
the bar increases from a M$_{Bmin}$ to a M$_{Bmax}$ value. To this end, we have
first chosen a model in which a low mass bar is already present. The morphology
of the POs   at a given energy, is determined by the presence of the radial and
vertical resonances \citep{spa02a}. Such resonances exist in any model of a
rotating 3D potential. Nevertheless, in the case of a model of a galactic bar,
the influence of the bar on the dynamics of the disc has to be conspicuous and
affect significantly the shape of the bar-supporting orbits. Therefore, in our
model A, we have chosen $GM_{D}=$0.87$, GM_{S}$=0.08 and $GM_{B}$=0.05, which
corresponds to a low mass bar. Following \citet{spa02a}, we have chosen
$\Omega_{b} =0.054$, which places the Lagrangian points L1 and L2 at a radius of
about 6.38~kpc.

\subsubsection{The stability diagram}
\label{mA_starpo}

A tool for describing the dynamics of 3D Hamiltonian systems is the stability
diagram \citep{cm85}. It describes the evolution of the linear stability of
families of POs, as one parameter (in our case \ej\!) varies and indicates the
critical energies, at which new bifurcating families are introduced in the
system \citep[for details see][\S 2.11]{gcobook}.

The linear stability of POs is calculated by means of the method of \citet{b69}.
Details about the algorithm  can be found e.g.~in \citet{cm85} \citep[see
also][]{S01}. Here, we only mention that the variation of the stability with \ej
is characterized by the variation of two stability indices, b1 and b2, one of
which refers to radial and the other to vertical perturbations. A PO is stable
(S) if both bi$\,\in (-2,2)$, with $i=1,2$. If one of the two stability indices
is $|$bi$| > 2$, then the orbit is characterized as simple unstable (U), while
if both indices are $|$bi$|> 2$ it is called double unstable (DU). Finally, if
all four eigenvalues of the monodromy matrix are complex numbers off the unit
circle, the stability indices cannot be defined and the PO is called complex
unstable $(\Delta)$.

\begin{figure}
\begin{center}
\resizebox{85mm}{!}{\includegraphics[angle=270]{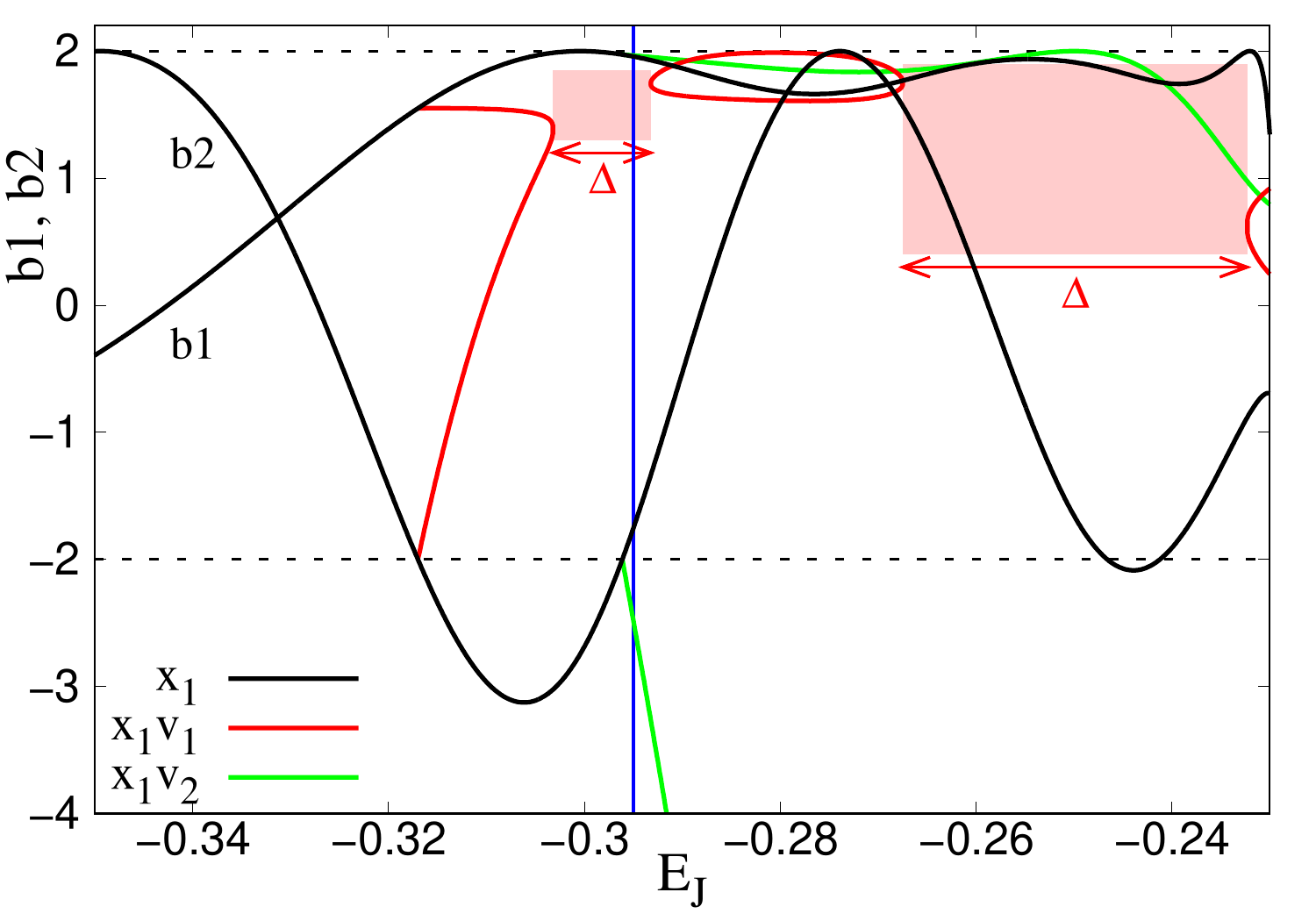}}
\end{center}
\caption{TI model A. Stability curves of the x1 (black curves, b1 and b2 are
respectively related to radial and vertical perturbations), x1v1 (red curves)
and x1v2 (green curves). The pink-shaded areas indicated by double edged arrows
and by the symbol $\Delta$, denote energy ($\mbox{E}_J$) intervals, where x1v1
is complex unstable. The horizontal dashed lines denote the critical values -2
and 2 of the stability indices, while the vertical blue line indicates the
energy value \ej=-0.295, at which we start our orbital calculations in the TD
models.}
\label{mA_stabdi}
\end{figure}

In Fig.~\ref{mA_stabdi} the stability curves of the x1 family are black, those
of x1v1 red and of x1v2 green. Index b1 is the one associated with the radial
and b2 with the vertical perturbations. The x1 family has a typical variation of
the stability indices \citep{spa02a}, with the b2 index having the standard
``S$\rightarrow$ U$\rightarrow$ S'' transition at the vILR region, bifurcating
x1v1 as S and x1v2 as U, at \ej $= -0.317$ and \ej $= -0.297$, respectively.
Beyond its bifurcating point, towards larger \ej\!\!, x1v1 has two successive
``S $\rightarrow \Delta \rightarrow$ S'' transitions. As we observe in
Fig.~\ref{mA_stabdi}, this has as a consequence the presence of two $\Delta$
intervals, for $-0.303<$\ej\!$<-0.293$ and $-0.267<$\ej\!$<-0.232$. The
variation of the stability indices in the autonomous case shown in
Fig.~\ref{mA_stabdi}, essentially determines the appropriate energy at which we
have to start our orbital explorations in the models with parameters varying in
time. In model A, we have chosen that energy to be \ej=$-0.295$ (denoted by the
vertical blue line), since at this value the families x1, x1v1 and x1v2 coexist.
Their representatives are S (x1), $\Delta$ (x1v1) and U (x1v2). These families
of simple-periodic orbits are considered to be the most important building
blocks of the boxy bulges in autonomous models \citep{psa02}.

\subsubsection{Navigation in Phase Space}
\label{mA_navi}

Periodic orbits determine the orbital content that supports observed
morphological features in dynamical systems like the TI galactic models, since
they determine the topology of the phase space. Around any stable PO exist
stability tori, while the presence of unstable PO introduces chaos \citep[see
e.g.][chapter 2]{gcobook}. In 2D systems, when the initial conditions (ICs) of a
PO are perturbed, we can directly observe on a 2D surface of section if the
displaced ICs belong to a stability island or to a chaotic zone. Then, by
integrating the displaced ICs of the PO, within a pre-defined time interval, we
also know if it is bar-supporting or not.

In 3D Hamiltonian systems, the 6D phase space can be reduced to a 4D space for
the ``surfaces'' of section and for the arrays of the ICs that uniquely
determine a PO \citep[see e.g.][]{spa02a}. Having the bar along the y-axis and
considering the $y=0$ plane as our surface of section, the coordinates of our 4D
spaces are $(x,z,p_x,p_z)$. We always will refer throughout the text to  ICs of
orbits in our system by giving the numerical values of their coordinates in this
array. The visualization of such 4D surfaces of section is not trivial. A method
that led to the association of specific structures in the neighbourhood of
stable  POs, as well as in the neighbourhood of the various kinds of unstable
POs encountered in 3D Hamiltonian systems, has been introduced by \citet{pz94}
and successfully applied in galactic models by \citet{kp11, kpc11} and
\citet{kpc13}. However, a global visualization of the entire phase space, in
which several regular and chaotic orbits coexist, has additional technical
difficulties, although attempts toward this goal have already been done
\citep{RLBK14,LROBK14,OLKR16}. Thus, unlike what happens in the case of the 2D
surfaces of section, in the 4D spaces of section it is not straightforward to
actually observe if a perturbation of the ICs of a stable, for instance, PO will
lead to an orbit on an invariant torus of the same orbit, to an orbit in a
chaotic sea, or even on an invariant torus of another stable PO. In this task
we are still essentially blind and only by experience one can follow some
interesting paths. As we will see below, in non-autonomous systems the
complexity of the situation increases.

In order to detect possible candidates of orbits that support the bar morphology
in TD models, we first study the phase space structure of the autonomous case
and use it as a basis in our further investigations. We use 2D, $(x,p_x)$,
surfaces of section on the equatorial plane around the main family x1, as well
as the $(z,p_z)$ projections of the 4D space of section, in which the orbits are
integrated for time $t = 5$~Gyr. This latter projection has been proven
especially helpful, because within a properly chosen integration time, the
resulting figure resembles a surface of section of a 2D case. This allows us to
trace directions, along which we can reach orbits in the ILR region that can be
used for building a rotating bar \citep[see e.g.~Figure 14 in][]{pk14a}.

\begin{figure*}
\begin{center}
\includegraphics[width=\textwidth]{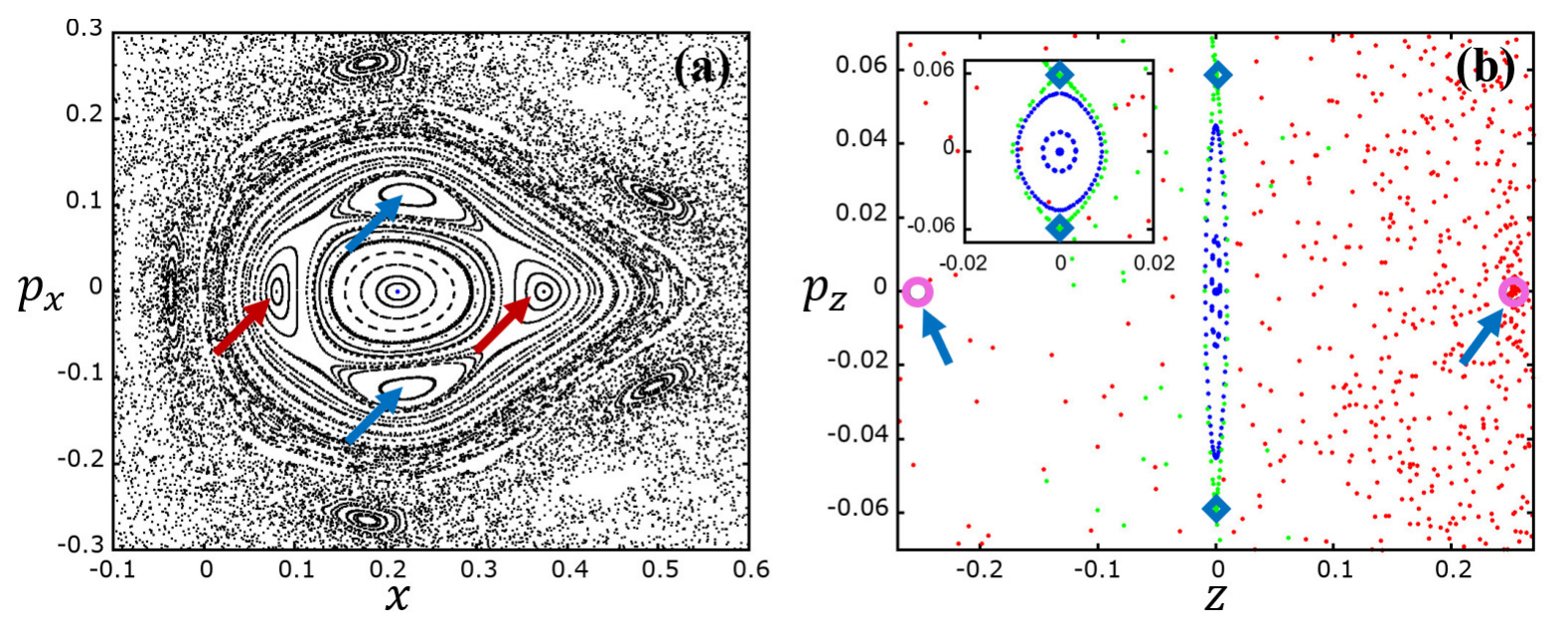}
\end{center}
\caption{TI model A for \ej=-0.295. (a) The $(x,p_x)$ surface of section around
x1, located at the center of the innermost stability island. Arrows point to the
four islands of the two POs of multiplicity 2, rm21 (red) and rm22 (blue) (b)
The $(z,p_z)$ projection of the Poincar\'{e} section around x1. Open circle
symbols indicate the locations of x1v1 (right) and x1v1$^{\prime}$ (left), while
the location of x1v2 ($p_z>0$)  and x1v2$^{\prime}$ ($p_z<0$) are indicated with
``diamonds''. Red points correspond to a perturbation of the x1v1 PO and they
are mainly concentrated in the right-hand side of the figure. The blue points
around x1 correspond to perturbed orbits lying on tori around x1 in the 4D
space, while the green points belong to a perturbation of the x1v2 PO (see also
embedded frame).}
\label{mA_soses}
\end{figure*}

For our ``starting-point'' in model A, at \ej=$-0.295$, such surfaces of section
are depicted in Fig.~\ref{mA_soses}. The usefulness of such diagrams in
identifying b/p-supporting orbits will become evident below. The main POs in the
two panels are x1 (S), with initial conditions $(x_0,z_0,p_{x_0},p_{z_0})\approx
(0.213,0,0,0)$, x1v1 $(\Delta)$ with initial conditions about
$(0.213,0.253,0,0)$ and x1v2 (U) with initial conditions about
$(0.211,0,0,0.059)$.

In the $(x,p_x)$ surface of section in Fig.~\ref{mA_soses}a, the x1 family,
being stable, is at the center of the main stability island (blue dot). The
central region around x1 is flanked by two sets of two islands (indicated with
red and blue arrows) of the known 2-periodic families rm21 and rm22 \citep[for
the origin of these families see][]{path19}.  For orbits on the equatorial
plane, surfaces of section like the one in Fig.~\ref{mA_soses}a allow us to find
out the morphology of any orbit displaced by $x$ or $p_x$ from x1, simply by
integrating its ICs.

In Fig.~\ref{mA_soses}b we give a $(z,p_z$) projection of the 4D surface of
section. It includes the main POs, as well as orbits in their neighbourhood and
outlines the basic structure of the phase space. Fig.~\ref{mA_soses} helps us
finding orbits that potentially sustain the 3D bar, as well as the dynamical
mechanisms that are in action. We observe that the stable x1 at (0,0) has the,
expected, tori around it. Orbits on these tori give the blue consequents along
elliptical curves around x1, resembling in this particular projection, the
invariant curves we encounter in 2D surfaces of section.

The two blue ``invariant curves'' belong to two orbits with  ICs
$(0.212...,0,0,0.015)$\footnote{We will indicate throughout the paper the ICs of
the perturbed orbits that remain identical to those of the corresponding  PO
\textit{truncated} at three decimal digits, followed by three dots. E.g.~in this
case, the first number in the array corresponds to the exact $x_0$ IC of the
x1v1 PO.} and $(0.212...,0,0,0.045)$. They are better viewed in the embedded
frame, in the upper left corner of Fig.~\ref{mA_soses}b, where the area around
x1  is depicted with a different scaling of the axes. The POs x1v1 ($z>0$) and
x1v1$^{\prime}$ ($z<0$), known in the relevant literature as ``frown''-``smile''
pairs, are located at the centers of the drawn open circle symbols and are also
denoted by arrows, close to the left and right sides of the frame of
Fig.~\ref{mA_soses}b, at $p_x=0$. A nearby to the $(\Delta)$ x1v1 orbit with ICs
(0.25,0.253...,0,0), i.e.~perturbed in the x-direction, is depicted with red
points. These consequents appear scattered mainly to the right of the x1
``stability island'' and only a few of them are observed in the left side.  The
consequents corresponding to integration time of around 1~Gyr, would appear even
more concentrated around the x1v1 initial conditions. The location of the U
representatives of the families  x1v2 ($p_z>0$) and x1v2$^{\prime}$ ($p_z<0$)
are indicated  with diamond ($\Diamond$) symbols. The plotted orbit at the
immediate neighbourhood of x1v2 (green points) has ICs $(0.215,0,0,0.059...)$.
Due to the proximity of the ICs of this orbit to those of the unstable periodic
one and to the last torus of x1, the green consequents stick initially to
the x1 ``island'' before they drift away into a chaotic sea.

\subsubsection{Orbital morphology and stability}
\label{mA_morstab}

\begin{figure}
\begin{center}
\includegraphics[width=1.0\columnwidth]{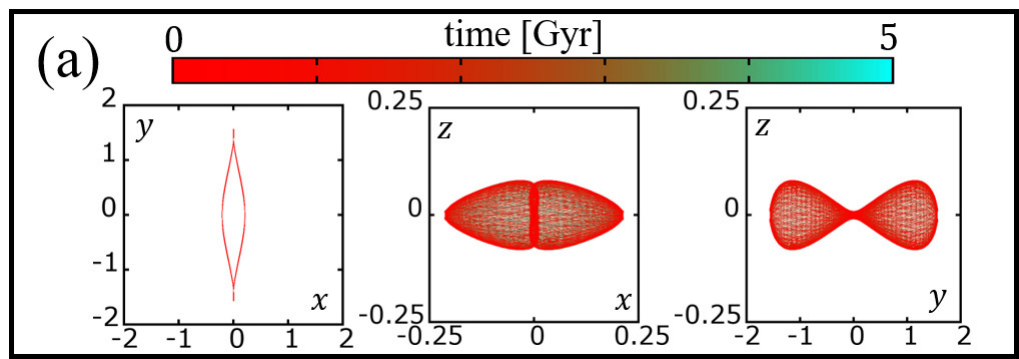}
\includegraphics[width=1.0\columnwidth]{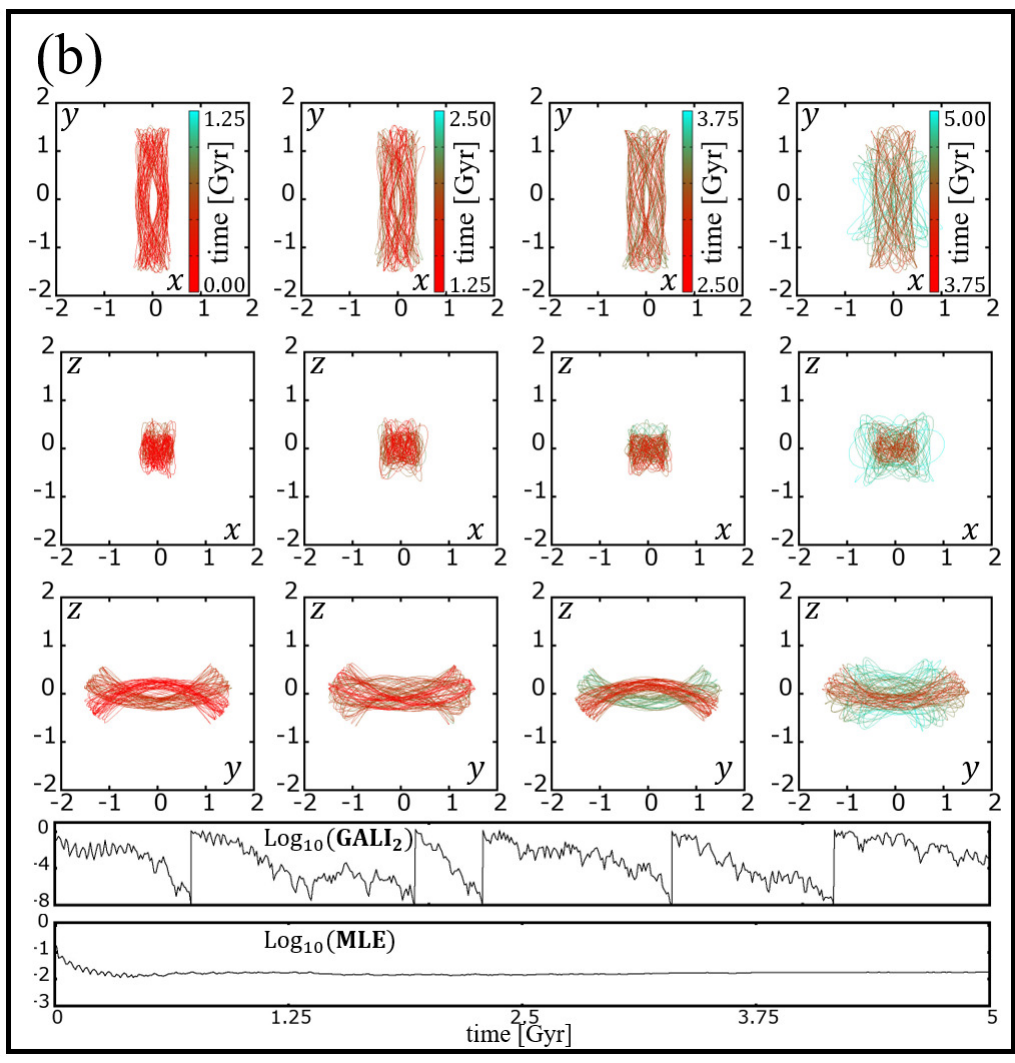}
\includegraphics[width=1.0\columnwidth]{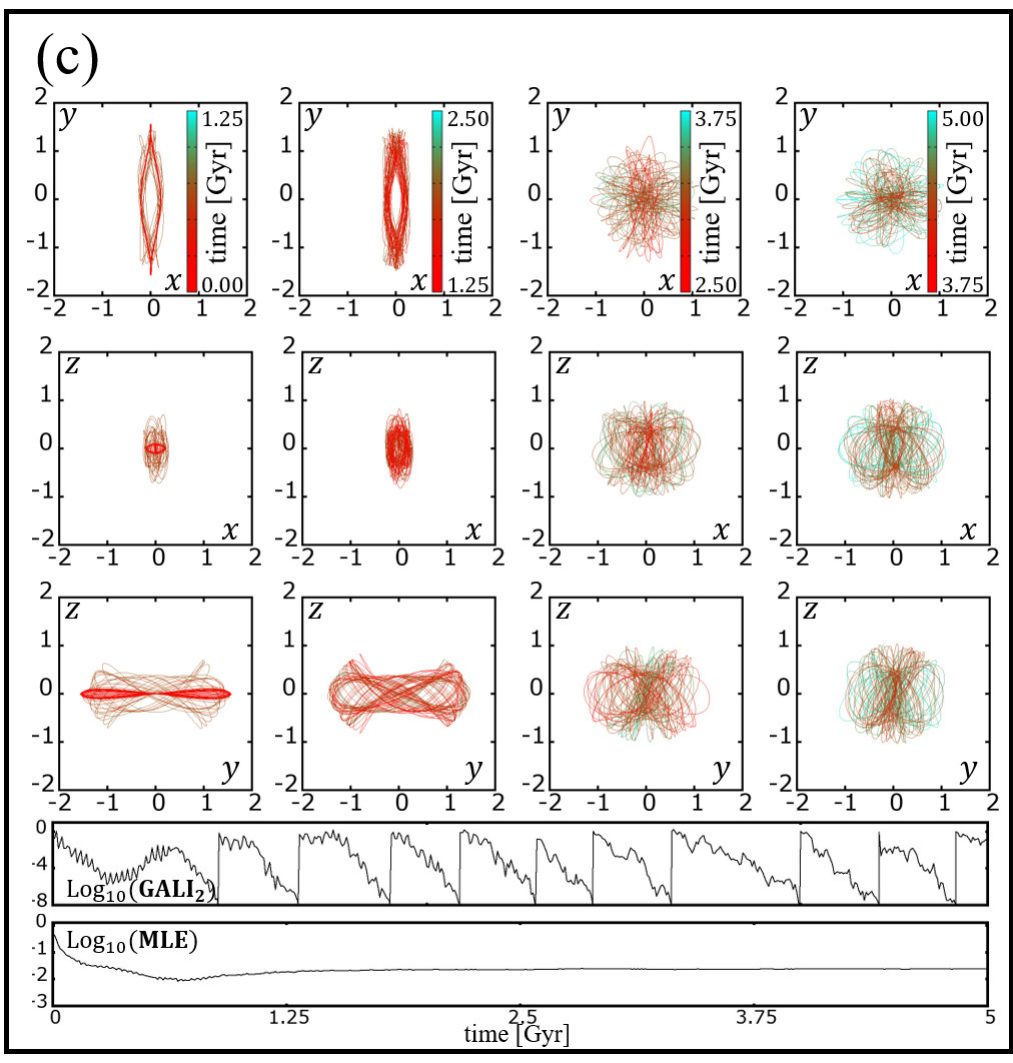}
\end{center}
\caption{TI model A. (a) The $(x,y)$, $(x,z)$ and (y,z) projections
(left, middle and right panel respectively) of a 3D regular orbit close to x1,
integrated for 5~Gyr. (b) The evolution of an orbit close to x1v1 $(\Delta)$. In
the upper three rows we see the evolution of the $(x,y)$, $(x,z)$,
$(y,z)$ projections for  times  (left to right columns) $0 \leq t \leq
1.25$~Gyr, $1.25<t \leq 2.5$~Gyr, $2.5<t \leq 3.75$~Gyr and $3.75<t \leq 5$~Gyr.
The variation of its GALI$_2$ and MLE is shown at the bottom, elongated panels.
(c) Similar to (b) but for an orbit close to x1v2 (U). The orbits are coloured
according to their evolution in time, as indicated by the related colour bars.}
\label{tigalis3a}
\end{figure}

We have to keep in mind the following features of
the main types of bar-supporting orbits, which  we calculated in the
autonomous model for the comparisons that will follow:
\begin{itemize}
\item \textbf{Orbits in the x1 neighborhood:}
The blue invariant-like curves around x1 in Fig.~\ref{mA_soses}b belong to
orbits we found by perturbing its $p_z$ IC. We note that their side-on
projections reinforce an ``$\infty$''-like morphology, which becomes more
striking as they approach the  ICs of x1v2 \citep{pk14a}. In a way, we can say
that these orbits  are associated morphologically with the U PO x1v2. A regular,
vertically perturbed x1 orbit, very close to x1v2, with ICs
$(0.212...,0,0,0.05)$, is given in Fig.~\ref{tigalis3a}a. The orbit does not
change essentially shape during the integration period. The three panels
correspond (from left to right) to projections in the  $(x,y)$,
$(x,z)$ and $(y,z)$ planes, which, with respect to the bar, correspond to the
face-on, end-on and side-on projections. The orbits in each individual window
are coloured according to time, from red (at the beginning of the integration)
to light blue (at the end of the integration), as indicated by the colour bar
above the three frames
\item \textbf{Orbits in the x1v1 neighborhood:}
We found orbits close to the $\Delta$ x1v1 PO, which reinforce a boxy side-on
view for considerably long times. We give in Fig.~\ref{tigalis3a}b a perturbed
by $\Delta x$, x1v1 orbit with ICs (0.25,0.253..,0,0). In this and all
subsequent similar figures, if not otherwise indicated, the orbits are presented
in three successive rows, which give the evolution from top to bottom, of the
face-on, end-on and side-on projections. The morphology of each orbit is
depicted in each row in four successive time windows; from left to right:  $0
\leq t \leq 1.25$~Gyr, $1.25<t \leq 2.5$~Gyr, $2.5<t \leq 3.75$~Gyr and $3.75<t
\leq 5$~Gyr. The colour bar indicating the evolution of the orbit in time, is
given in this and all subsequent similar figures, at the right-hand side of each
panel.

The orbit in Fig.~\ref{tigalis3a}b, has a boxy shape in its face-on projection
(upper row) and reinforces a boxy morphology in the side-on view for 3.75~Gyr
(three first panels of the third row). For $t>3.75$~Gyr, the enhancement of
boxiness is less evident. The variation of GALI$_2$ and MLE in the two
elongated, panels at the bottom of  Fig.~\ref{tigalis3a}b, indicates the weakly
chaotic nature of the orbit. We observe that its  GALI$_2$ requires relatively
long time intervals to reach very small values (GALI$_2 \leq 10^{-8}$), after
which the reinitialization process described in Section \ref{sec:intro} is
implemented. Also, the MLE tends to saturate towards the end of the integration
to a positive value with $\log_{10}\mbox{MLE} \approx -1.74$.

By considering, besides the variation of the indices, the morphological
evolution of the orbit, we conclude that it is a sticky chaotic one
\citep{ch08}. This is in agreement with \citet{cppsm17}, who claim that sticky
orbits with boxy projections both in the face-on and side-on views are usually
sticky chaotic. By trying several ICs in the neighbourhood of this $\Delta$
x1v1 PO, we realize that there is a broad sticky region
surrounding x1v1.

\item \textbf{Orbits in the x1v2 neighborhood:}
A perturbed by $\Delta x$, x1v2, U, orbit, with ICs (0.215,0,0,0.059...),
reinforces for 2.5~Gyr the bar and the b/p bulge (Fig.~\ref{tigalis3a}c). Its
side-on projection during the first 1.25~Gyr (third row), after a period being
$\infty$-shaped, puffs up, following a boxy, hybrid morphology between x1v2 and
x1v1. Then, for $1.25<t \leq 2.5$~Gyr it keeps this shape, having mainly a
x1v2-like side-on morphology. For larger time its chaoticity becomes more
evident. The chaotic nature of this orbits is also reflected in the evolution of
its GALI$_2$, which experience many reinitializations during the integration
time, and its MLE, which attains a positive final value $\log_{10}\mbox{MLE}
\approx -1.61$. Although the evolution of the MLE of the orbits in
Figs.~\ref{tigalis3a}b and \ref{tigalis3a}c tells us that the overall behavior
of both orbits is chaotic (actually the evolutions of the two MLEs are very
similar) it fails to vividly depict the dynamical differences of the two orbits.
On the other hand, the frequent reinitializations of GALI$_2$ for the orbit of
Fig.~\ref{tigalis3a}c clearly indicate its higher degree of chaoticity. This
advantage of the GALI$_2$ method will become even more significant in the case
of TD systems, where orbits can experience epochs of regular or chaotic
behaviors during their evolution, because the MLE is not adequate to follow
subtle changes in the dynamics \citep{mbs13}. For these reasons we prefer to use
GALI$_2$ as chaos indicator for the orbits studied in this paper. Nevertheless,
the MLE has been calculated for all orbits presented in the paper (being always
in agreement with GALI$_2$ for the overall behavor of orbits), although we do
not give its variation in the subsequent figures.
\end{itemize}
The orbits described above represent in no case all kinds of non-periodic orbits
one may encounter in the phase space of our system at the given \ej\!\!. They
are some of the orbits, which have been found in previous studies to shape the
bar out of the equatorial plane. We will follow next the evolution of such
orbits as the mass of the bar increases with time, in order to check if they
retain or not their bar-supporting character. In this way we will compare the
evolution of the same  ICs in the TD case with those in TI models.
\subsubsection{Successive autonomous models}
\label{mA_manyauto}

\begin{figure}
\begin{center}
\resizebox{90mm}{!}{\includegraphics[angle=270]{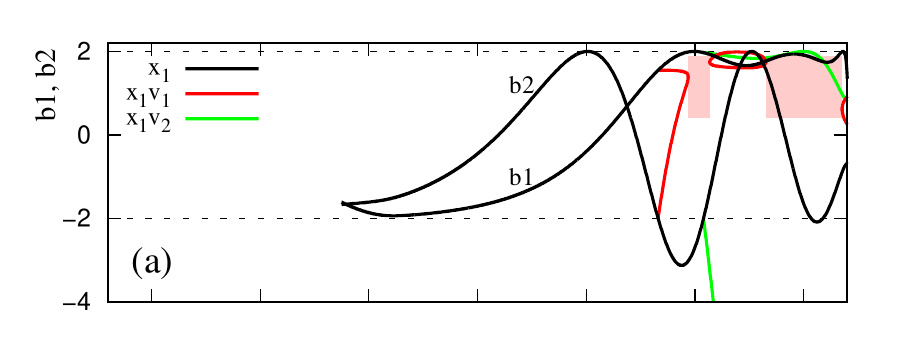}}
\resizebox{90mm}{!}{\includegraphics[angle=270]{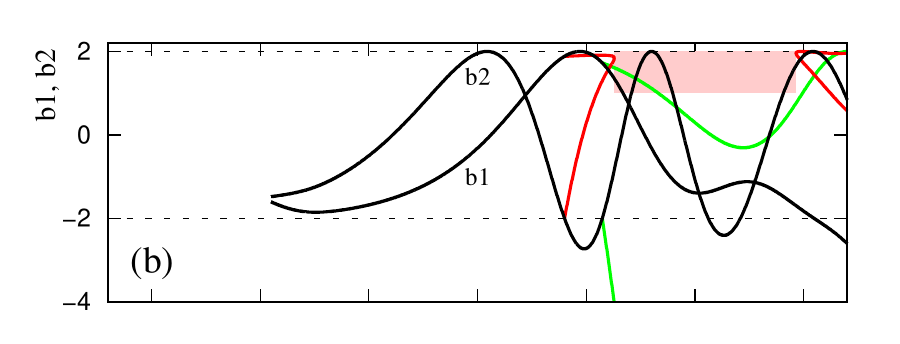}}
\resizebox{90mm}{!}{\includegraphics[angle=270]{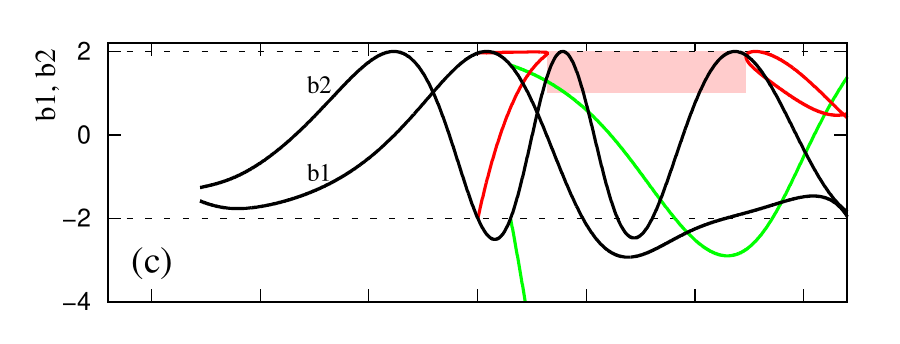}}
\resizebox{90mm}{!}{\includegraphics[angle=270]{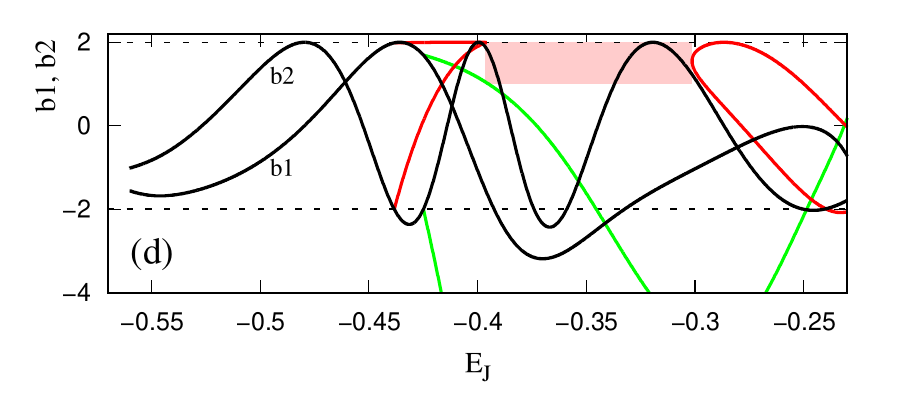}}
\end{center}
\caption{TI model A. Stability diagrams similar to the one of
Fig.~\ref{mA_stabdi} for (a) $GM_B$=0.05, (b) $GM_B$=0.1, (c) $GM_B$=0.15 and
(d) $GM_B$=0.2. The shaded pink areas indicate energy intervals where x1v1 is
complex unstable. The successive TI models with increasing $GM_B$ constitute the
shadow evolution of a TD model in which $GM_B$ increases by the same amount
within a predefined time.}
\label{shad_autostab}
\end{figure}

A possible way of studying the evolution of a TD system is to rely on a sequence
of potentials/models obtained from a sequence of snapshots from an $N$-body
simulation \citep[see e.g. the review by][]{ath13}. Here we follow a similar
route, i.e. we use a sequence of TI models, which we call the ``shadow
evolution'' of the corresponding TD model. These TI models have successively
increasing $M_B$ (or equivalently $GM_B$) between its minimum and maximum
values. For example, in Fig.~\ref{shad_autostab}, we present the variation of
the stability indices in a series of TI models, starting with the TI model A, in
which $GM_B$ increases successively from 0.05 to 0.2, i.e.~it quadruples. This
series of TI models constitutes the shadow evolution of a TD model in which
$M_B$ increases by the same amount within a predefined time interval.

The immediate information we obtain from the stability diagrams of
Fig.~\ref{shad_autostab}, is whether or not a PO of a given family, at a certain
\ej\!\!, retains its stability as $GM_B$ varies. In addition, for any model in
the range $0.05 \leq GM_B \leq 0.2$, we can calculate the dynamics in the
neighbourhood of any PO and compare it with the dynamics in the corresponding
region of the TD model, when it reaches the same $GM_B$ during its evolution.
Obviously, at the moment the orbit we evolve in the TD case reaches the $GM_B$
value we are interested in, it will have a different \ej than the one at the
beginning of its integration. Thus, we want to compare its morphology with that
of the orbit at the corresponding \ej of the autonomous model, as predicted by
its location in the Poincar\'{e} surface of section at that energy. We will
refer also to individual models in this series, as ``shadow'' models.
\subsection{The non-autonomous case}
\label{mA_nonauto}

In order to study the evolution of an orbit in a TD model, we start integrating
an initial condition at the chosen initial energy. For model A, with
$GM_B$=0.05, this is \ej=$-0.295$, because at this energy x1, x1v1 and x1v2
coexist. In all studied cases of the present work the quantity $GM_B$ increases
linearly from the minimum to its maximum value.

In practice, it is not computationally feasible to examine the evolution of all
orbits existing in a model as $GM_B$ increases.  Moreover, we do not know a
priori which orbits will be the important, bar-supporting ones in non-autonomous
models. One can realize this by trying to navigate him/herself in a 4D space of
section. For this reason, we selected characteristic cases of orbits, which have
been already found to play an important role in supporting the thick part of
rotating bars in autonomous models \citep{spa02a, psa02, pk14a, pk14b}. We start
from the main families of POs in the ILR region (x1, x1v1, x1v2)  and we apply
perturbations along directions chosen merely by our experience. In this effort,
for model A, we have used as compass Figs.~\ref{mA_soses}a,b. In particular, we
investigate the following cases:

\subsubsection{FAST $GM_B$  INCREASE}
\label{mA_fmb}

In the first TD model, the quantity $GM_B$ increases from 0.05 to 0.2,
i.e. it quadruples within 5~Gyr. At the same time, $GM_D$ decreases so that the
masses of the three components at each time step continue to satisfy the
condition $G(M_{D}+M_{S}+M_{B})=1$. The initial energy in all examined cases for
model A, is \ej=$-0.295$, as it was in the TI case, discussed in
Section~\ref{mA_autoc}.
\paragraph{Evolution of ICs of POs in the TD model:}
\label{mA_po}

\begin{enumerate}
\item  The evolution of x1: The first PO to be investigated is x1, on the
equatorial plane of the galaxy. The x1 family is considered as the most
important one in bar galaxy models. The x1 representative in Fig.~\ref{mA_soses}
has ICs $(0.212...,0,0,0)$. Its morphological evolution is given in
Fig.~\ref{mA_x1_fast}. In Fig.~\ref{mA_x1_fast}a we give it within the time
interval 0-3.75~Gyr, while in Fig.~\ref{mA_x1_fast}b in the period
$t$=3.75-5~Gyr. For this orbit, the corresponding GALI$_2$ index is presented in
Fig.~\ref{mA_x1_fast}c. Remarkably, the shape of x1 persists for more than
3.75~Gyr (Fig.~\ref{mA_x1_fast}a and initial phase in Fig.~\ref{mA_x1_fast}b)
and then it turns to a shape reminiscent of a bifurcation of x1 at the radial
3:1 resonance. As we know, the planar POs bifurcated at the radial 3:1 resonance
are introduced in the system in pairs. The one with a morphology reminiscent of
the one during the late stage evolution of the orbit in Fig.~\ref{mA_x1_fast}b
has two representatives, symmetric with respect to the y-axis \citep[see
e.g.][]{spa02a}. Thus, the presence of its both representatives in the
$t$=3.75-5~Gyr time interval would support locally boxy isodensities. The
evolution of the GALI$_2$ index (Fig.~\ref{mA_x1_fast}c) shows that the orbit we
examine practically retains its regular nature during its 5~Gyr evolution
despite its morphological change. We note, that GALI$_2$ does not register the
orbit as chaotic when its  morphological transformation occurs at
$t\gtrapprox$3.75, since the orbit before and after that time remains regular.
An indication of this transition between different regular behaviours, is the
rather
abrupt change of the inclination of GALI$_2$ at $t\approx$3.8. If for larger
times, we consider the trace of the orbit on the Poincar\'{e} surfaces of
section in the corresponding TI, i.e. in the shadow, models, we find that they
are located on  islands of stable 3:1 POs.
\begin{figure}
\begin{center}
\includegraphics[width=1.0\columnwidth]{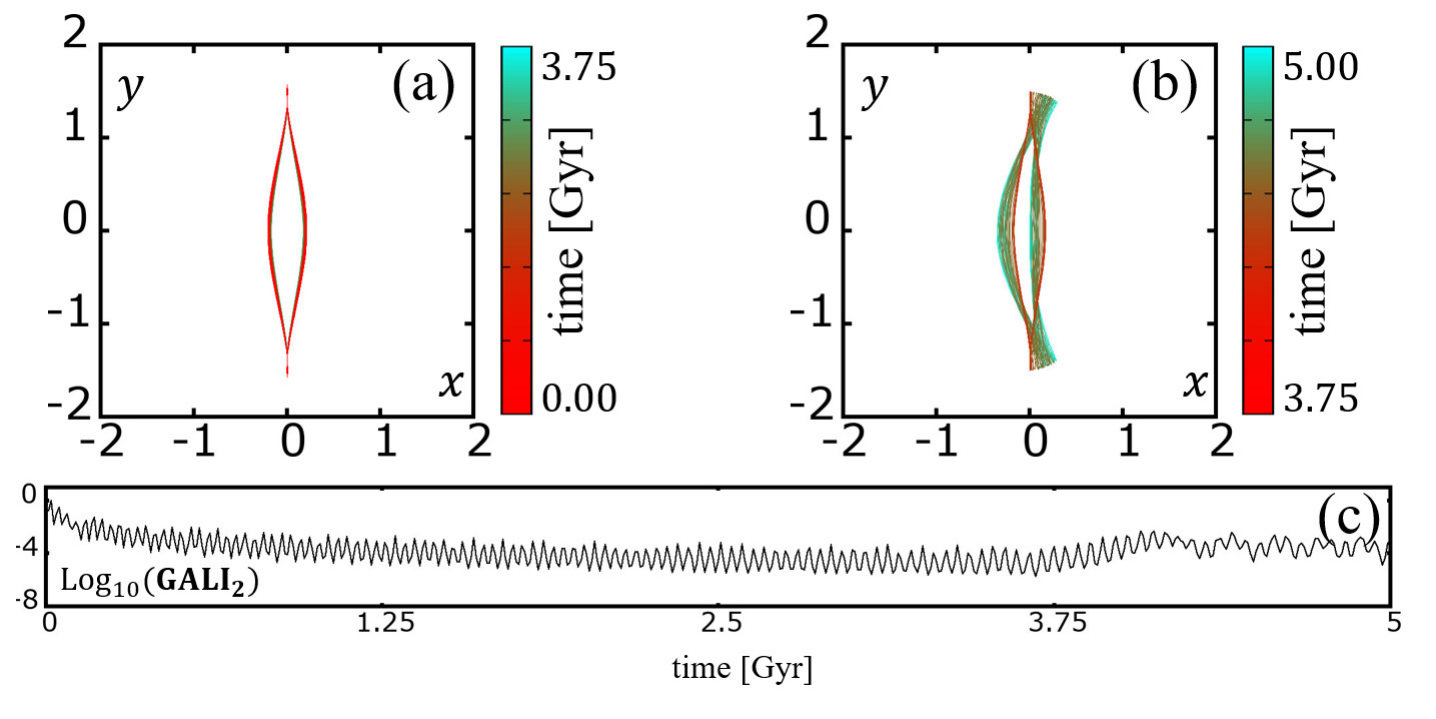}
\end{center}
\caption{Fast growing bar model A. Morphological evolution of x1 in the TD model
(see text for details). In (a) during $t=$0-3.75~Gyr and in (b) during
$t=$3.75-5~Gyr. The orbit is coloured according to time, as indicated by the
colour bars to the right-hand sides of the panels. In (c) we give the time
evolution of the GALI$_2$ index of the orbit.}
\label{mA_x1_fast}
\end{figure}
\item  The evolution of x1v1: We evolved the ICs $(0.212...,0.253...,0,0)$ of
the x1v1 PO of the autonomous case, which for \ej=$-0.295$ is $\Delta$. Now,
the exact shape of the orbit varies from the beginning of the integration.
Nevertheless, as we can observe in Fig.~\ref{mA_x1v1_fast}, it retains some
degree of regularity in its morphological evolution as it shows a clear
bar-supporting character, both in its face-on and edge-on views, especially at
the initial stages of its evolution. There is a  gradual change of its relative
dimensions by increasing its extent along the minor axis, something that becomes
conspicuous in the last integration period ($3.75<t\leq5$~Gyr) where the orbit
is chaotic as the evolution of GALI$_2$ (Fig.~\ref{mA_x1v1_fast}d) indicates.
Note that the orbit has an initial  quasi-regular behaviour for $t \lessapprox
1.4$, as its GALI$_2$ did not reach the threshold $\log_{10}\mbox{GALI}_2 \leq
10^{-8}$, which is followed by a chaotic epoch.
\begin{figure}
\begin{center}
\includegraphics[width=1.0\columnwidth]{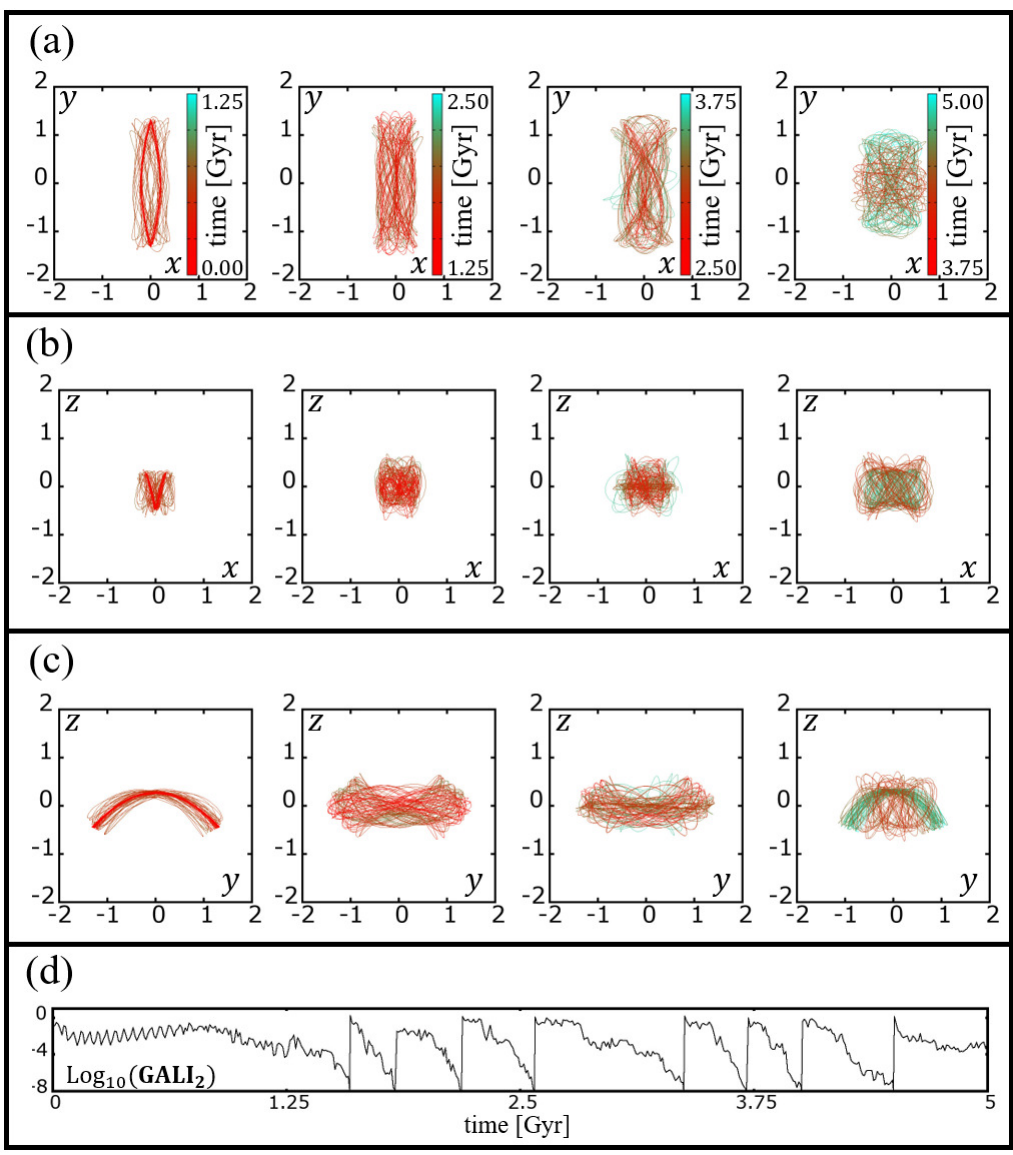}
\end{center}
\caption{Fast growing bar model A. The evolution, of the x1v1 PO in the TD
system. Projections in the (x,y) (a), (x,z) (b), and (y,z) (c) planes, for
times  (left to right) $0 \leq t \leq 1.25$~Gyr, $1.25<t \leq 2.5$~Gyr, $2.5<t
\leq 3.75$~Gyr, $3.75<t \leq 5$~Gyr. The orbit in all projection panels is
coloured according to time, as indicated by the colour bars in (a). (d) The
time evolution of the GALI$_2$ index of the orbit.}
\label{mA_x1v1_fast}
\end{figure}
The orbital shapes we find in the projections of the orbits in
Figs.~\ref{mA_x1v1_fast}a,b and c are typical of non-periodic orbits encountered
in the ILR region of similar autonomous systems. The face-on projections
(Fig.~\ref{mA_x1v1_fast}a) have initially a morphological transition from a
x1-type shape to a shape encountered in face-on projections of orbits sticky to
x1v1, supporting a face-on X feature \citep[cf with figure 7 in][]{pk14b} in the
window $0\leq t\leq 1.25$~Gyr (first panel of Fig.~\ref{mA_x1v1_fast}a), lasting
also for $1.25<t\leq 2.5$~Gyr (second panel). Then, during the next  time
interval, $2.5<t \leq 3.75$~Gyr (third panel), the morphology of the orbit is
similar to that of the multiplicity 3 PO rm33 \citep[see Table 4 in][]{path19}.
Finally, for  $3.75<t \leq 5$~Gyr (fourth panel) we observe again the appearance
of a face-on X feature, this time in a more squarish orbit. Here again we come
across morphologies found in the autonomous models. In \citet{pk14b} such orbits
are found by relatively large  perturbations in the $z$-direction (see figure
10d in that paper). In the first 3.75~Gyr the  orbit supports a bar with similar
length as the x1  PO in the region, while in the last time window, the orbit
shrinks in the direction of the major axis of the bar, gaining width along the
minor axis.

The edge-on views of the orbit (Figs.~\ref{mA_x1v1_fast}b,c) reinforce a
frown-smile morphology continuously during the 5~Gyr time interval. The
prevailing side-on shapes are close to frowns (first and last panel in
Fig.~\ref{mA_x1v1_fast}c), or smile-like (third panel), or hybrids (second
panel), resembling the shapes of peanut-supporting orbits in the neighbourhood
of $\Delta$ x1v1  POs presented in \citet{pk14a}.

The morphological evolution and the variation of the GALI$_2$ index, resembles
that of Fig.~\ref{tigalis3a}b, with the orbit being slightly more chaotic as
more reinitializations of GALI$_2$ are observed in Fig.~\ref{mA_x1v1_fast}d,
showing similar behaviours  to chaotic orbits in the neighbourhood of x1v1 of
the autonomous system. Actually, the orbit is always located in weakly chaotic
zones around the x1v1 ICs of the corresponding ``shadow'', autonomous models.
The simultaneous boxiness in the face-on (Fig.~\ref{mA_x1v1_fast}a), edge-on
(Fig.~\ref{mA_x1v1_fast}b) and side-on (Fig.~\ref{mA_x1v1_fast}c) projections is
in agreement with the result of \citet{pk14b} and \citet{cppsm17} about the
double boxiness of sticky chaotic orbits in rotating bars.
\item  The evolution of x1v2: From the results we present in
Fig.~\ref{mA_x1v2_fast} it is evident that the U PO x1v2, with ICs
$(0.210...,0,0,0.0591...)$ at \ej=$-0.295$ in the autonomous case, has in its
face-on view a morphological evolution similar to x1 during the 5~Gyr
integration period. Namely, its face-on view (Fig.~\ref{mA_x1v2_fast}a) presents
the same transformation to a 3:1-like bifurcation of x1, as x1 does after
3.75~Gyr  (Figs.~\ref{mA_x1_fast}a,b). The orbit in its side-on projection
(Fig.~\ref{mA_x1v2_fast}b) is always restricted inside the known $\infty$-shape
of the outline of the x1v2 POs in this projection. Notably, the height of the
orbit is reduced with time and after 5~Gyr the resulting shape is almost planar.
This evolution points to a regularly behaving orbit and this is confirmed by the
evolution of the GALI$_2$ index (Fig.~\ref{mA_x1v2_fast}c).
\begin{figure}
\begin{center}
\includegraphics[width=1.0\columnwidth]{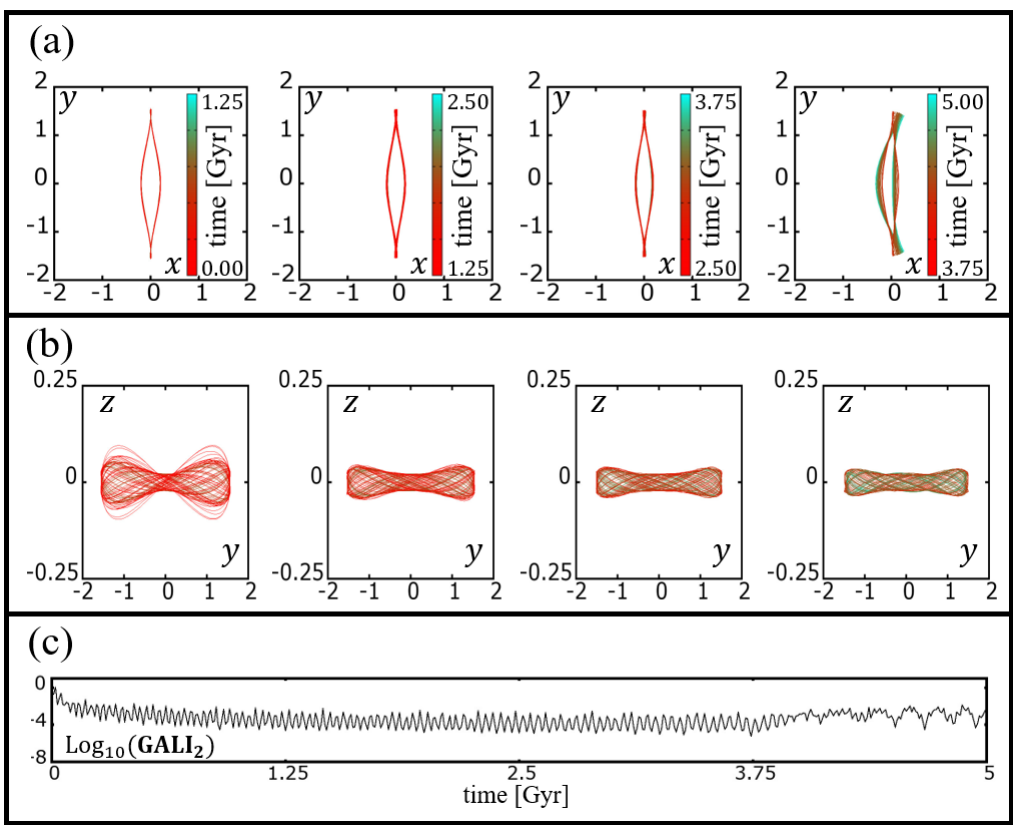}
\end{center}
\caption{Fast growing bar model A. The evolution, in the TD system, of the ICs
of the x1v2 PO.  The (x,y) (a) and (y,z) (b) projections  for times  (left to
right) $0 \leq t \leq 1.25$~Gyr, $1.25<t \leq 2.5$~Gyr, $2.5<t \leq 3.75$~Gyr,
$3.75<t \leq 5$~Gyr. Note in (b) the  different scale in the axes. The orbit in
all projection panels is coloured according to time, as indicated by the colour
bars in (a). (c) The time evolution of the GALI$_2$ index of the orbit.}
\label{mA_x1v2_fast}
\end{figure}
This result sounds counter-intuitive, since a U PO of the autonomous case
behaves as regular when it evolves in the TD model. Nevertheless, this is not a
property of these particular ICs, as also other x1v2 PO of the TI model we
considered for different \ej values,  show similar behaviours when they are
evolved in the TD system. In order to understand this behaviour we checked the
position of the evolved orbit in the phase space of the autonomous models of the
shadow evolution taken at times in the middle of the four periods we plot our
orbits throughout the paper (namely at $t=0.625$, 1.875, 3.125 and 4.375~Gyr,
for which we respectively have $GM_B=0.06875$, 0.10625, 0.14374, 0.18125 and
\ej=0.306, 0.328, 0.350, 0.372). In particular, we considered the locations of
the $(x,z,p_x,p_z)$ coordinates of our orbit in the TD system by registering its
upwards intersections with the $y=0$ plane for each time window. By placing
these ``traces'' of the orbit on the $(x,p_x)$ surfaces of section, as well as
on the $(z,p_z)$ projections of the shadow models, we realize that they
correspond to points belonging to x1 tori. This is seen in Fig.~\ref{empty1}
where these points are plotted with heavy red dots. It is clear that all red
points in the first three time windows of Fig.~\ref{empty1}a are very close to
the initial conditions of x1 at the corresponding autonomous models. In the last
period x1 appears unstable in the TI model (it is located between the two main
stability islands on the $p_x=0$ axis) and the red points of the time-dependent
orbit drift towards the stability islands of one of the bifurcated 3:1 orbits.
The regular behavior of the orbit is also seen in its projections in the
$(z,p_z)$ plane (Fig.~\ref{empty1}b) as its points always form a `ring'
distribution around the planar x1 orbit (denoted by the blue dot at $z=p_z=0$.)
Thus, the morphological evolution of our orbit is fully in agreement with the
dynamical behavior predicted by the models in the shadow evolution described in
Fig.~\ref{shad_autostab}.
\begin{figure*}
\begin{center}
\includegraphics[width=\textwidth]{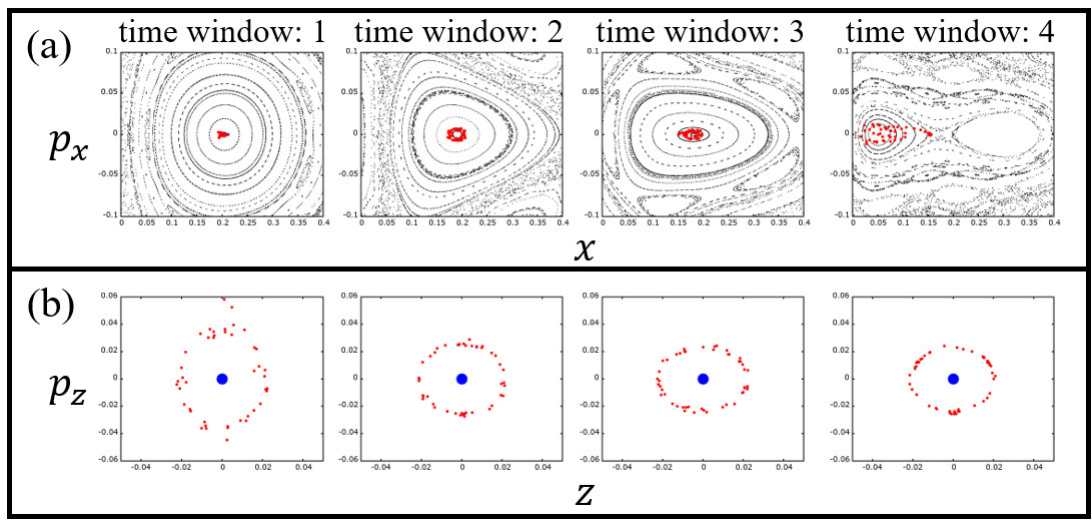}
\end{center}
\caption{Fast growing bar model A. (a) The
    $(x,p_x)$ surfaces of section of the autonomous, ``shadow'' models
    at times (from left to right) $t=0.625$, 1.875, 3.125 and 4.375~Gyr
    (\ej and $GM_B$ values of the shadow models are given in the
    text). Heavy red dots indicate the ``traces'' of the orbit,
    starting with the ICs of the x1v2 PO of the TI model at
    \ej=$-0.295$, as it evolves in the TD system within each time
    window, 1: $0 \leq t \leq 1.25$~Gyr, 2: $1.25<t \leq 2.5$~Gyr, 3:
    $2.5<t \leq 3.75$~Gyr, 4: $3.75<t \leq 5$~Gyr. (b) The $(z,p_z)$
    projection of the same orbit at the same time windows. The blue
    dot indicates the position of the x1 PO at $z=p_z=0$.}
\label{empty1}
\end{figure*}
\end{enumerate}

\paragraph{Evolution of ICs of non-POs of the TI model in
the TD system:}
\label{mA_nonpo}
Despite the fact that POs are the backbones of any barred model, the orbital
content of real bars consists of non-POs (regular or chaotic). In studies of TI
models, perturbations along certain directions have been proven particularly
interesting for supporting observed structures, such as the peanut-shaped part
of the bars. So, we examined the evolution of initial conditions along these
directions in the TD model A:

\begin{enumerate}
\item Perturbations of x1: Firstly, we applied radial and vertical perturbations
to the x1 representative of the autonomous model at \ej$=-0.295$. The radial x1
perturbations led in general to regular, bar-supporting, orbits with
morphologies known from studies of TI models. Typical evolutions of perturbed x1
orbits are given in Fig.~\ref{mA_x1pert_rad}.  In Figs.~\ref{mA_x1pert_rad}a,b
the perturbed x1 orbits with ICs (0.212...,0,0.05,0) and (0.212,0,0.1,0)
respectively, remain bar-supporting during the whole period of the 5~Gyr.
However, especially in Fig.~\ref{mA_x1pert_rad}a, we observe a morphological
transformation with time, associated with a weakly chaotic epoch during this
transition, as the GALI$_2$ variation below the orbits indicates. Nevertheless,
the orbit can always be characterized as bar-supporting. Similar morphologies
are encountered when we start integrating in this TD model, initial conditions
on the invariant curves in the neighbourhood of x1 of our TI model A
(Fig.~\ref{mA_soses}a).
\begin{figure}
\begin{center}
\includegraphics[width=\columnwidth]{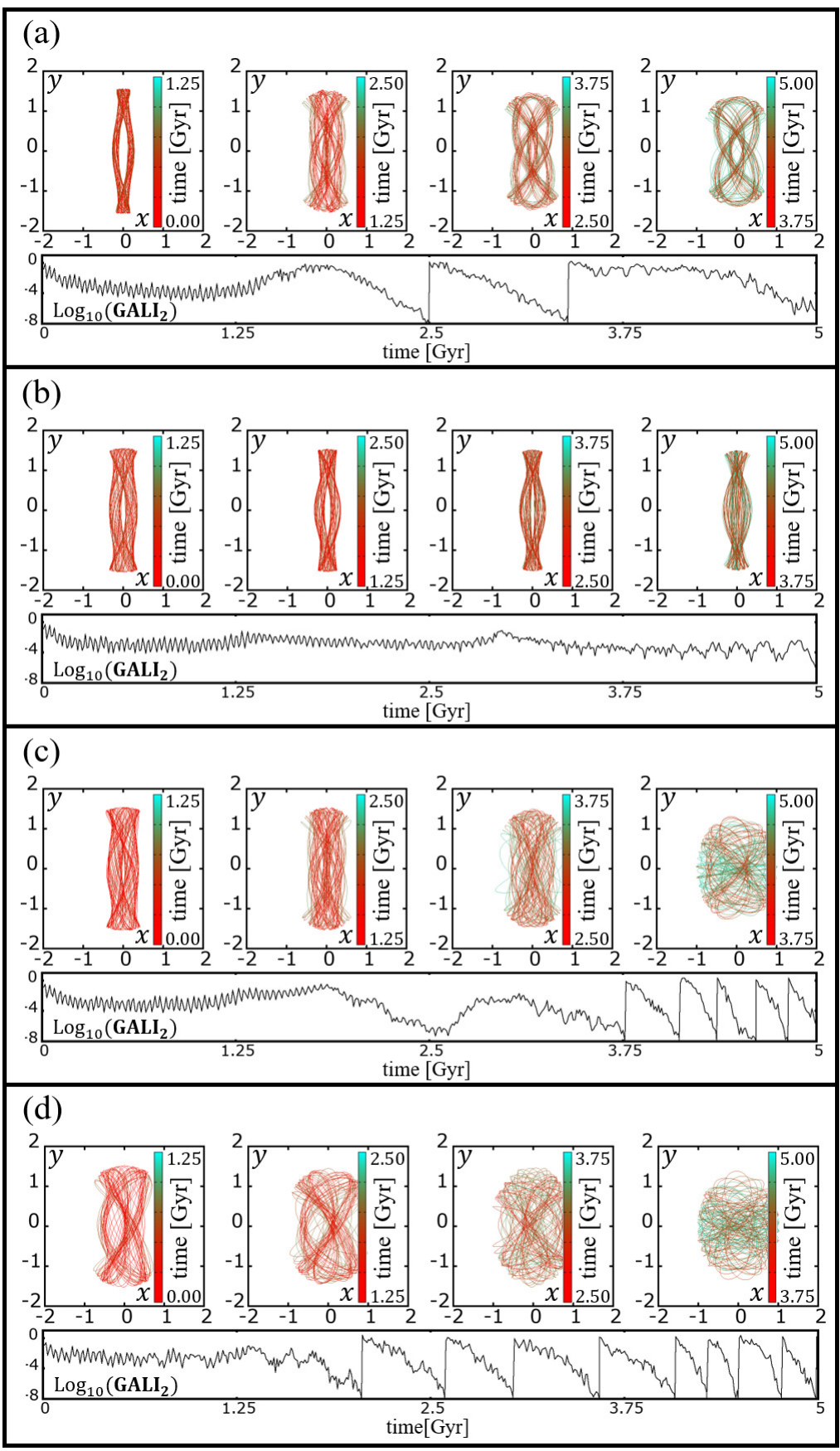}
\end{center}
\caption{Fast growing bar model A. The evolution of morphology and GALI$_2$
index of typical radially perturbed x1 orbits. The ICs of the orbits in (a) and
(b) are on invariant tori around x1 in the autonomous case, while in (c) and (d)
are located close to the edge of the x1 stability island, where chaos and tiny
stability islands are present.
Time windows and the colours of each orbit are as in Fig.~\ref{mA_x1v1_fast}.}
\label{mA_x1pert_rad}
\end{figure}

Morphologically, the evolution in Fig.~\ref{mA_x1pert_rad}a,  leads
from a x1- to a rm33-like \citep{path19} shape. The rm33 shape has
been frequently encountered in the evolution of bar-supporting orbits
in TD models in our study \citep[see also figure 9
  in][]{mm14}. Another frequently encountered evolution of perturbed
x1 orbits is given in Fig.~\ref{mA_x1pert_rad}b. This is characterized
by the prevalence of the ansae-type morphology as time
increases. Effectively, this is similar to the transformation of the
x1-like shape to a ``double'' 3:1 orbit
(see also the discussion about boxiness due to 3:1 bifurcations of x1
in paragraph \ref{mA_po}).

Contrarily to the orbits in Figs.~\ref{mA_x1pert_rad}a,b, when we start
integrating orbits located at the edge of the x1 stability island of the TI
model, in a region dominated by the presence of tiny stability islands and
chaotic zones (Fig.~\ref{mA_soses}), we find only partly bar-supporting orbits.
Two examples are given in Figs.~\ref{mA_x1pert_rad}c,d with ICs
(0.212..,0,0.15,0) and (0.212..,0,0.2,0) respectively. Their GALI$_2$ variation,
after an initial period of regular behaviour points to moderate chaoticity,
especially during the last time windows. The orbit's evolution for $2.5<t\leq
3.75$~Gyr in Figs.~\ref{mA_x1pert_rad}c is associated once again with the
appearance of an X-feature in the face-on view of the orbit
\citep{pk14b,tp15,cppsm17}. We also note that the orbit in
Figs.~\ref{mA_x1pert_rad}d, supports during its regular phase in the first
2.5~Gyr an rm21-like \citep{path19} morphology.

Small vertical x1 perturbations lead to ICs belonging to one of the invariant
tori surrounding the PO in Fig.~\ref{mA_soses}b.
Qualitatively, their side-on projections remain morphologically invariant during
the 5~Gyr period. In Fig.~\ref{mA_x1pert_zdot} we give the side-on profiles of
an orbit, whose ICs are (0.212...,0,0, 0.05), during the first (a), and last
(b), time windows. During the whole time of integration (5~Gyr), such orbits
remain very close to the equatorial plane, thus for presenting their
$\infty$-like shapes in this projection, the scales on the axes are not equal.
We note though, that the height they reach is reduced as time increases, being
minimum in the time interval 3.75-5~Gyr. In parallel, the face-on projections
remain close to a x1 morphology up to $t=3.75$ and then we have again the usual
in this model transformation to a 3:1-like morphology, similar to the one
described in Figs.~\ref{mA_x1_fast} and \ref{mA_x1v2_fast}. We found similar
evolution for all x1 orbits perturbed in the $p_z$ coordinate with $0.01\leq
p_{z_0} \leq 0.6$. The GALI$_2$ variation is the characteristic one for regular
orbits, so we avoid giving it here.
\begin{figure}
\begin{center}
\includegraphics[width=\columnwidth]{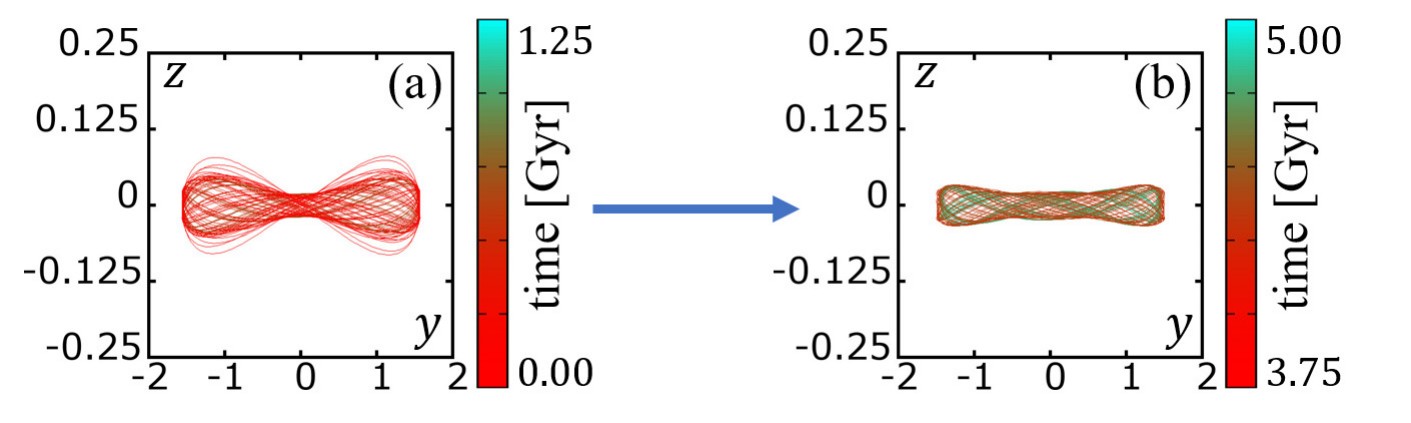}
\end{center}
\caption{Fast growing bar model A. The side-on view, in the TD system, of a
perturbed by $p_z$ x1 orbit, starting on a torus surrounding the PO in the TI
model. (a) The orbit during $0 \leq t \leq1.25$~Gyr and (b) during  $3.75<t \leq
5$~Gyr. Note the different scales in the axes.}
\label{mA_x1pert_zdot}
\end{figure}

At this point, before proceeding with investigating further the behaviour of
non-POs in our TD potential, we would like to underline the following: Even if
we increase the $p_z$ perturbation to reach the immediate neighbourhood of x1v2
in Fig.~\ref{mA_soses}b the regular behaviour of the orbits in the TD potential
persists. Starting with ICs (0.212...,0,0, 0.06) (cf.~embedded frame in
Fig.~\ref{mA_soses}b) we observe again the regular behaviour seen in
Fig.~\ref{mA_x1pert_zdot}.
This happens despite the fact that the orbit with the same
 ICs, (0.212...,0,0,0.06) in the \textit{TI}
  potential has a chaotic behaviour during the 5~Gyr integration
period. In Fig.~\ref{mA_TI_x1closex1v2} we give the evolution of the
face-on (a), side-on (b) and the GALI$_2$ variation for this
orbit. Evidently, it has a strong bar-supporting character only during
the first 1.25~Gyr and consequently a totally different morphological evolution
in its side-on profile compared with the orbit starting with the same ICs in
the TD model (Fig.~\ref{mA_x1pert_zdot}).
\begin{figure}
\begin{center}
\includegraphics[width=\columnwidth]{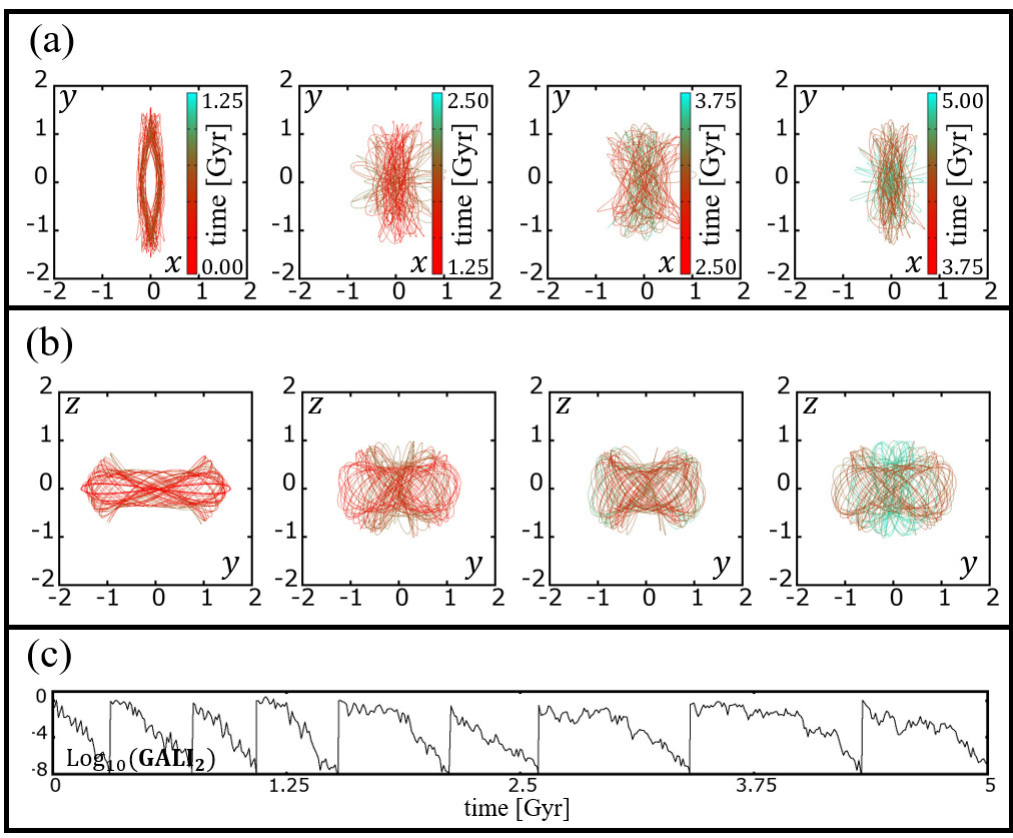}
\end{center}
\caption{\textit{TI} model A. Plots similar to
    Fig.~\ref{mA_x1v2_fast}, but for an orbit with ICs
    (0.212...,0,0, 0.06), in the chaotic sea around x1v2.
The same ICs integrated in the fast growing bar model A, give a
regular orbit with morphologies similar to that of the orbits
presented in Fig.~\ref{mA_x1_fast} (face-on) and
Fig.~\ref{mA_x1pert_zdot} (side-on).}
\label{mA_TI_x1closex1v2}
\end{figure}

In a way we observe here a mechanism that organizes a  chaotic orbit by adding
time-dependency in the system. However, we can understand this behaviour by
following the location of the orbit we integrate on the $(x,p_x)$ and $(z,p_z)$
projections of successive TI models in our shadow evolution. As $GM_B$
increases, the orbit moves closer to x1, in the stability region occupied by the
invariant tori around x1 (Fig.~\ref{mA_soses}b) and follows morphological
patterns similar to those expected by integrating orbits on these x1 tori of the
autonomous case. The traces of the orbit have a vary similar distribution on the
surfaces of section of the autonomous models as those in Fig.~\ref{empty1}.

We move on now to the evolution of ICs of characteristic non-POs, in the
neighbourhood of the two other main  POs existing at this energy in the
corresponding TI model.

\item Perturbed x1v1 orbits:
In the autonomous model A, at \ej=$-0.295$, we have a x1v1 $\Delta$
representative. Thus, there are no quasiperiodic orbits around it. We
have perturbed radially and vertically its initial conditions and have
followed its evolution in the time-dependent case.

Firstly, we examined the radial x1v1 perturbations. In all applied
radial perturbations in the $x$-direction, i.e.~for $0.25\leq x_0 +
\Delta x \leq 0.37$, where $x_0\approx 0.213$, all examined orbits
were bar-supporting during the first 1.25~Gyr. For larger times, the
supported bar structures were gradually dissolved. However, larger
deviations from the $x_0$ coordinate of x1v1, do not necessarily lead
to a faster drift into chaos.  In
  Figs.~\ref{mAfx1v1r}a,b we respectively see  the evolution of the
  face-on and side-on views of a characteristic orbit of this type,
  having ICs  (0.37,0.253...,0,0). In this particular case, we have a
deviation from the $x_0$ initial condition of x1v1 $|\Delta x_0| =
0.157$  and the orbit is now partially bar-supporting for $t\lessapprox
2.5$~Gyr. The evolution of its GALI$_2$ index
  (Fig.~\ref{mAfx1v1r}c) indicates a chaotic character,
  which becomes more pronounced  as soon as the morphology of the
  orbits ceases being bar-supporting.
\begin{figure}
\begin{center}
\includegraphics[width=\columnwidth]{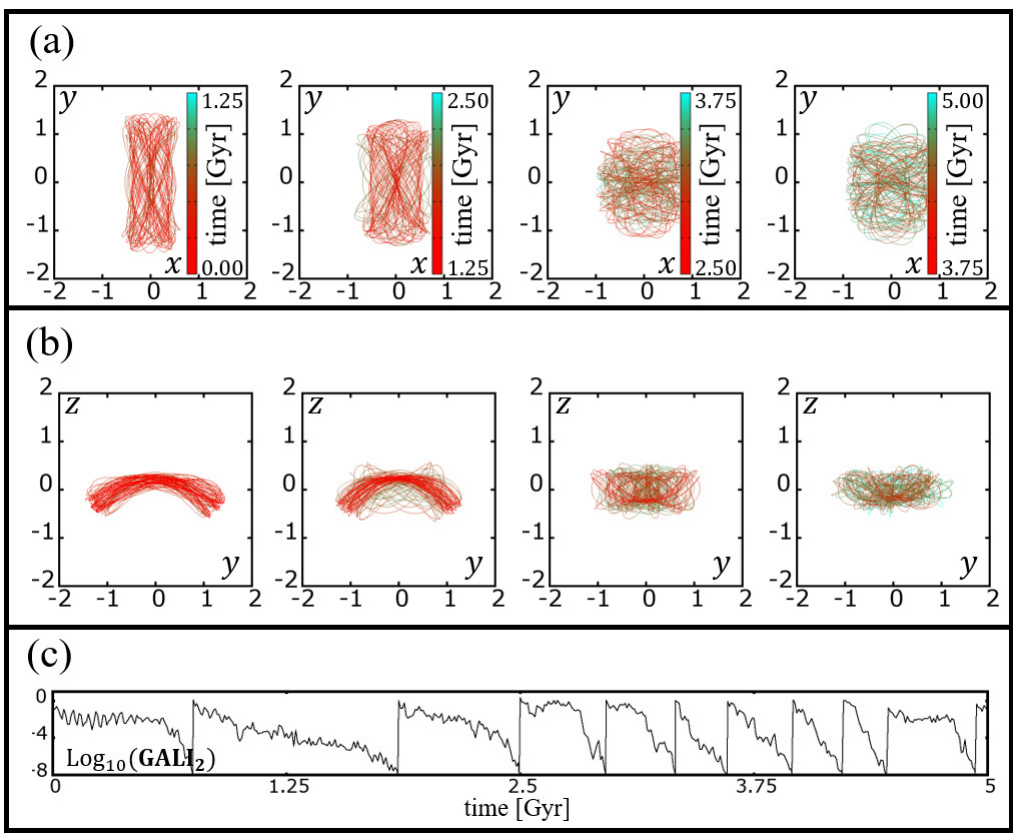}
\end{center}
\caption{Fast growing bar model A. Plots similar to
    Fig.~\ref{mA_x1v2_fast} but for a characteristic bar-supporting
    orbit for $t\lessapprox 2.5$~Gyr, obtained as a radial
    perturbation of the x1v1 PO of the TI system.}
\label{mAfx1v1r}
\end{figure}

As regards vertically perturbed orbits in the neighbourhood of x1v1,
most perturbations we applied in the $z$-direction, led to chaotic
orbits. Nevertheless, in the neighbourhood of the
 PO $(0.18 \leq z_0 \leq 0.3,$ with $z_0$(x1v1)
$\approx 0.253)$ the orbits are partially bar supporting, mainly
during the first 1.25~Gyr.

\item Perturbed x1v2 orbits:
By applying perturbations to the ICs of the U PO x1v2, we reach regions of
phase space essentially already discussed in the presentation of the evolution
of perturbed x1 orbits above. Radially perturbed x1v2 POs with
$0.205<x_0<0.215$, led consistently to regular orbits similar to what we have
presented in Fig.~\ref{mA_x1pert_zdot} with a frequent transition of the x1-like
face-on shape to a 3:1-like one at times $t>3.5$~Gyr as in the case of
Figs.~\ref{mA_x1_fast} and \ref{mA_x1v2_fast}.
Vertical perturbations of the U x1v2 PO behave also in this case like regular
orbits for long time intervals. This happens either because they are initially
located on the x1 tori we observe in the $(z,p_z)$ projection of the surface of
section in the autonomous case (better seen in the enlarged frame in
Fig.~\ref{mA_soses}) and continue behaving as such in the TD evolution, or
because there is a shifting of the traces of the orbits with respect to the
surfaces of section of the corresponding autonomous models in the shadow
evolution, as the one described in Fig.~\ref{empty1}.

\end{enumerate}

\subsubsection{SLOW $GM_B$ INCREASE }
\label{mA_smb}

Motivated by the growth of M$_B$ in the $N$-body simulation by
\citet[][see their Figure 2]{mm14} we studied as well the evolution of
characteristic orbits in models with a slow increase of the mass of
the bar. There are long time intervals in $N$-body
simulations, during which we have only a moderate or slow increase of
M$_B$. Thus, starting with the same initial conditions of the orbits
for \ej=$-0.295$, we studied the evolution of characteristic orbits
when, during 5~Gyr, we have a growth of the mass of the bar from GM$_B
= 0.05$ to GM$_B = 0.06$, i.e. when we have only a 20\% increment.

\paragraph{Evolution of ICs of POs in the TD model:}
\label{mA_spo}

\begin{enumerate}
\item The evolution of x1: The  orbit with ICs the
  ones of the x1 PO of the TI model, remains practically unchanged
  during the 5~Gyr period (and for this reason we do not show it), as
  if it has been evolved  in the TI model. It does not exhibit a
  transition to a 3:1-like morphology for $t>3.5$~Gyr, as in the
  previous, ``fast'' growing bar case (Fig.~\ref{mA_x1_fast}).
\item  The evolution of x1v1: The evolution of x1v1 in the
  slow-growing bar model within 5~Gyr, can be described in general as
  similar to the one in the fast $GM_B$ growth case up to 3.75~Gyr
   (Fig.~\ref{mA_x1v1_fast}), and for this reason we
    do not plot it. We did not encounter the final shrinking of the
  orbit along the major axis in the late stage of its evolution,
  evident in the right panels in Fig.~\ref{mA_x1v1_fast}. Contrarily,
  despite being $\Delta$ in the autonomous case, when the bar is
  slowly growing, the orbit keeps practically its periodic
    orbital shape during the first time
    window ($ t \leq 1.25$~Gyr). However, for larger times, it
    starts following a ``thick-elliptical'', quasiperiodic-like
morphology ($1.25<t \leq 2.5$~Gyr), and later it develops a
    weakly-chaotic ($2.5<t\leq 3.75$~Gyr) and eventually a chaotic
    character ($t>3.75$).

\item The evolution of x1v2: The orbit has a x1v2-like character with narrow
$\infty$-type side-on profiles, as in the ``fast'' case (Fig.~\ref{tigalis3a}a)
without even having the x1- to 3:1-like transition.  The evolution of this orbit
is again remarkable, given that its ICs correspond to a U PO in the autonomous
model.
\end{enumerate}

\paragraph{Evolution of ICs of non-POs of the TI model in the TD
system:}
The typical orbital dynamics in the TD system for
  orbits with ICs in the neighbourhood of the three main families of
  POs of the TI model can be summarized as follows:

\begin{enumerate}
 \item  Perturbations of x1: Radial  perturbations of x1 lead to regular orbits
with morphologies either similar to quasiperiodic orbits around x1, or with
morphologies similar to  POs of a higher multiplicity, existing at the x1
neighbourhood in the beginning of the integration. In the slowly growing bar
model A, we do not find any further morphological evolution as in the radially
perturbed x1 orbits of the ``fast case'' (Fig.~\ref{mA_x1pert_rad}). The orbits
retain their initial  shapes. Vertical perturbations of x1 in its immediate
neighbourhood lead to the usual orbits with an envelope of $\infty$-like shape
in their side-on views, for the cases with $0.01\leq p_z \leq 0.06$, we
examined. Again here, the height of the orbits in the side-on projections is
reduced with time.

\item Perturbations of x1v1: Firstly, we evolved
   orbits with ICs close to the $\Delta$
   PO x1v1 of the autonomous case for \ej=$-0.295$,
  by varying its $x_0$ coordinate. As in the autonomous and in the
  fast growing bar case, smaller perturbations in the $x$-direction
  were not necessarily associated with more regular, or more
  bar-supporting orbits. The evolution of the orbits in this model
   is, in most cases, similar to the evolution of
  the orbits with the same  ICs  in the TI model.
  Thus, there are orbits which retain a frown or smile,
  quasiperiodic-like, side-on projection during the whole 5~Gyr
  period.

Similar morphologies are encountered also in x1v1 orbits vertically perturbed
along the $z$-direction. We find again ICs that seem to belong to a volume of
phase space of the autonomous model, which, when  evolved in the slowly bar
growing  model A, support the bar, however, this time for $t \lessapprox
2.5$~Gyr. To avoid redundancy we do not plot such orbits as they exhibit the
same known patterns namely a quasiperiodic-x1-like face on view shape, combined
with a ``thick'' frown, or smile, one in the side-on projections.

Finally, an interesting class of morphological patterns encountered
usually for $t\gtrapprox2.5$~Gyr is presented in
Fig.~\ref{mAs_x1v1_z}. The  depicted, particular
  orbit has ICs (0.212...,0.18,0,0) and for
 $t\lessapprox2.5$~Gyr retains a bar-supporting
morphology with boxy face-on and a frown-smile side-on
shapes. Nevertheless, we present it here for its morphology during the
last part of its integration  (two last panels of
  Figs.~\ref{mAs_x1v1_z}a,b,c). We also note that its GALI$_2$
  indicator (Fig.~\ref{mAs_x1v1_z}d) points to a moderate chaotic
  orbit during the whole evolution. In the last two time
 windows of Fig.~\ref{mAs_x1v1_z}a its face-on views
  are round and its extents along the minor and major axes are almost
equal  (last panels in Fig.~\ref{mAs_x1v1_z}b and
  Fig.~\ref{mAs_x1v1_z}c for respectively the end-on- and the side-on
  views). Their overall size is comparable with that of the b/p part
of the bar in our model. We want to underline at this point the
appearance of a kind of X-shaped structure that can be observed in the
edge-on views of orbits like this (both end-on and side-on). The
sharpness of the X features and the angles of their wings make it
rather unlikely to be combined with the X structures supported by
orbits associated with x1v1 or x1v2 in a unique
morphology.
\begin{figure}
\begin{center}
\includegraphics[width=\columnwidth]{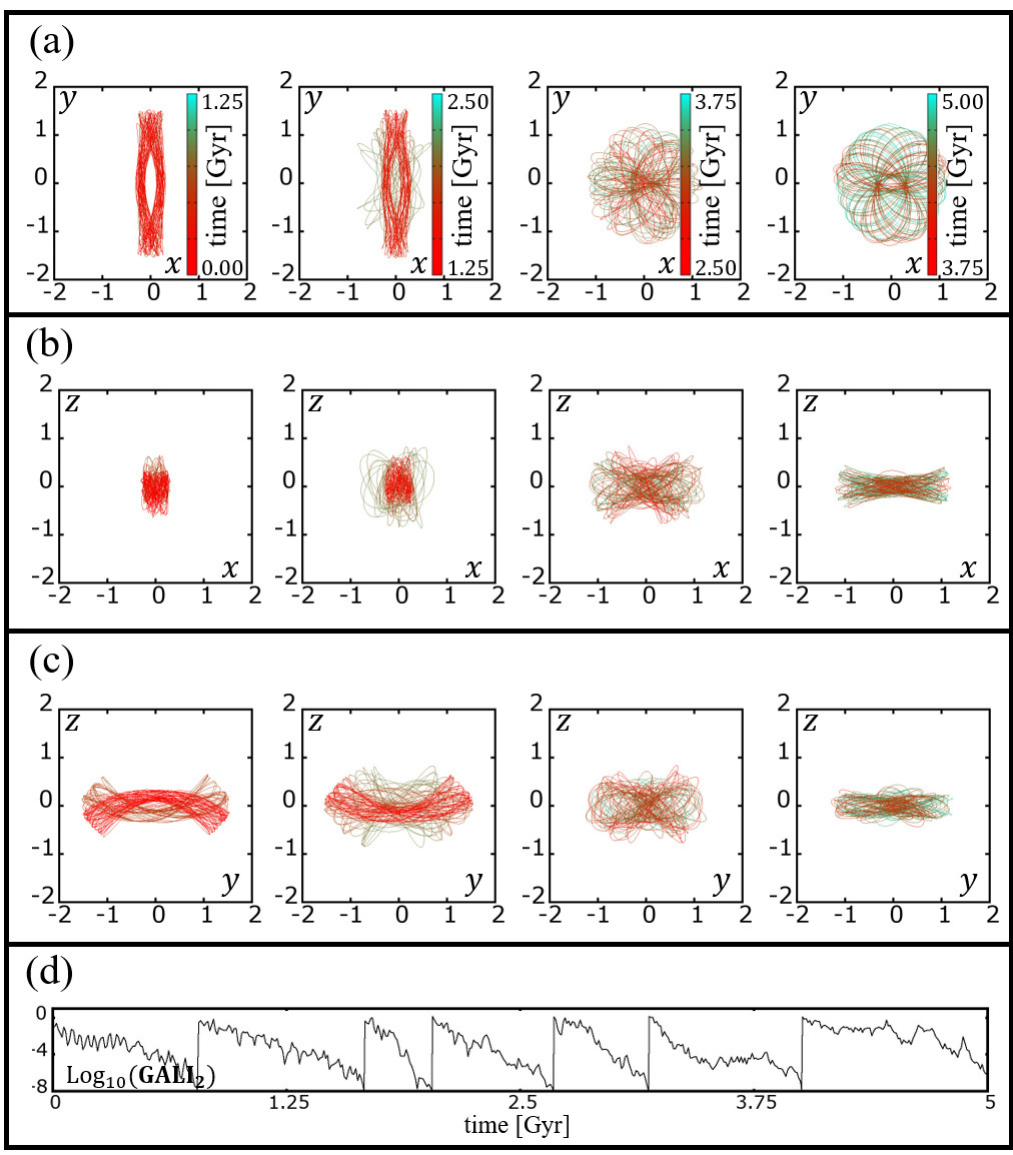}
\end{center}
\caption{Slowly growing bar model A. Plots similar to Fig.~\ref{mA_x1v2_fast}
but for a vertically perturbed x1v1 $\Delta$ orbit with a bar-supporting
evolution for $t\lessapprox2.5$~Gyr, which develops an X feature in the edge-on
views for $t\gtrapprox2.5$~Gyr, while the bar has been dissolved.}
\label{mAs_x1v1_z}
\end{figure}

\item Perturbations of x1v2: Radially perturbed x1v2 orbits, in the immediate
neighbourhood of the PO $(0.205\leq x_0(\rm{x1v2})\approx 0.21 \leq 0.215)$,
remain morphologically invariant as in all previous models we examined. They
follow the usual shape with a x1v2-like envelope (Fig.~\ref{tigalis3a}a). The
same holds for x1v2 orbits perturbed in the $p_z$-direction, with $|p_z| <
|p_{z_0}\rm{(x1v2)}|$ as in the autonomous case (Fig.~\ref{mA_soses}b). For
$|p_z|> |p_{z_0}\rm{(x1v2)}|$ we find a zone with orbits that morphologically
can be described as having hybrid x1v1-x1v2 morphologies, like e.g. the orbit
with ICs (0.210...,0,0,0.63) (not shown in the paper). Despite the fact that in
the autonomous model x1v1 is $\Delta$, this purturbed orbit behaves like a
sticky one, trapped for a certain time around x1v1 tori of stable x1v1 POs
\citep[cf with figure 13 in][]{pk14a}.
\end{enumerate}

\section{Model B: A massive bar}
\label{mB}

Our model B is one with a massive bar already present at the beginning  of its
evolution, having $GM_{D}=$0.79, $GM_{S}$=0.08 and $GM_{B}$=0.13, which
corresponds to a bar 2.6 times as massive as the bar of model A.

\subsection{The autonomous case} \label{mB_autoc}

\begin{figure*}
\begin{center}
\includegraphics[width=\textwidth]{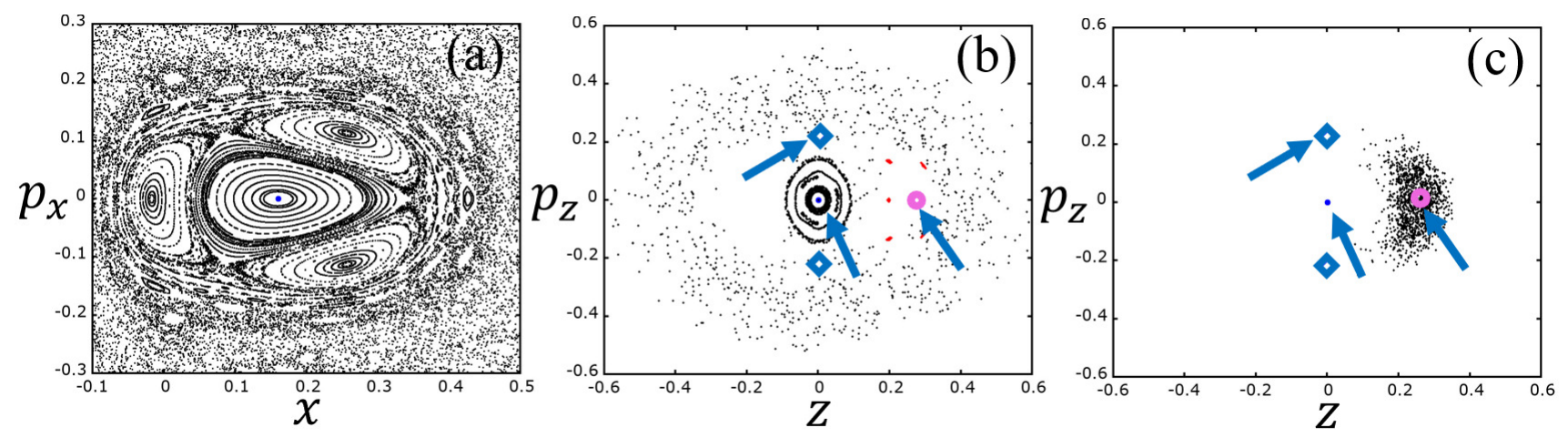}
\end{center}
\caption{TI model B for \ej=$-0.352$. (a) The
    $(x,p_x)$ surface of section around x1 (located at the center of
    the innermost stability island and indicated by a blue dot). (b)
    and (c) The $(z,p_z)$ projection of the Poincar\'{e} section
    around x1 (blue point at $z=p_z=0$, which is also indicated by an
    arrow). The open circle symbol (accompanied by an arrow) indicates
    the location of x1v1, while the location of x1v2 ($p_z>0$) and
    x1v2$^{\prime}$ ($p_z<0$) are indicated with ``diamonds'' (with an
    arrow also showing the position of x1v2). Note in (b) the presence
    of the five tori (coloured in red) of a quasiperiodic orbit
    surrounding the $\Delta$ x1v1 PO. Black points in (b) correspond
    to vertical perturbations of x1, while in (c) they belong  to
    several orbits with ICs in the neighbourhood of the x1v1
    PO.}
\label{mBsoses}
\end{figure*}

Qualitatively, the evolution of the stability of the central family x1 is
similar to the one in model A (comparable to the one in
Fig.~\ref{shad_autostab}c). As energy varies, the two other main families, x1v1
and x1v2 are introduced in the system at the vILR of the model, at a standard
S$\rightarrow$U$\rightarrow$S transition of x1. The x1v1 family is introduced as
S, but soon after its bifurcation it becomes $\Delta$, changing back to S
at a larger \ej\!\!. The other 3D, 2:1 family, x1v2, is introduced as U and
remains as such. We have chosen at the starting point the energy \ej= $-0.352$,
in which x1v2, bifurcated from x1 at \ej=$-0.368$, has already b2$\approx-7.7$.
Meanwhile x1v1 is complex unstable ($\Delta$) for E$_J\gtrapprox-0.356$. The ICs
of the main POs are for x1 (S) $(x_0,z_0,p_{x_0},p_{z_0})\approx (0.16,0,0,0)$,
for x1v1 $(\Delta)$ about (0.152,0.277,0,0) and for x1v2 (U) about
(0.142,0,0,0.223).  The surfaces of section, which will help us in understanding
the differences between the TI and TD cases for model B are depicted in
Fig.~\ref{mBsoses}.

\begin{itemize}
\item  \textbf{Orbits in the x1 neighbourhood:}  The shape of the radially
perturbed x1 orbits are the predicted by their location in the $(x,p_x)$ surface
of section (Fig.~\ref{mBsoses}a). A significant difference with respect to model
A, is that in model B, at \ej$=-0.352$, there are no rm21/rm22 orbits, as the
absence of the characteristic four islands around x1 in Fig.~\ref{mBsoses}a
indicates. However, a major zone around the central x1 region is occupied by the
stability islands of a 3-periodic orbit belonging to the rm33 family
\citep{path19}. Evidently, this affects the shape of the radially perturbed x1
orbits.

Vertical perturbations of x1 (either in the $z-$ or $p_z-$direction) lead to
x1v2-type morphologies, with $\infty$-like envelopes in their side-on
projections, as in Fig.~\ref{tigalis3a}a. This happens as long as the ICs of the
orbit are projected inside the area located in Fig.~\ref{mBsoses}b between the
ICs of x1v2 and x1v2$^{\prime}$ (diamond points), which is occupied by tori
around x1.  Beyond this region, we find weakly chaotic and chaotic orbits,
determined by the structure of the phase space in the neighbourhood of the two
other main families, x1v1 and x1v2 (see below).

\item  \textbf{Orbits in the x1v1 neighbourhood:}
Perturbed x1v1 orbits, in the $x$-direction, in the interval $0.13
\leq x_0 \leq 0.22$ (with $x_0$(x1v1) = 0.1517...), have consistently
a regular character, reflecting a quasiperiodic-like morphology that
could be vaguely described as a ``thick'' x1v1 PO.
For slightly larger perturbations, e.g. for $x_0$ = 0.25, we still
have a frown side-on profile, combined with boxy, face-on projections
that harbour an X feature, which is typical of sticky orbits
\citep{pk14a, cppsm17}.

Similar is the orbital dynamics of the vertically perturbed x1v1
orbits in the $z$-direction, with $z_0 < z_0$(x1v1)$\approx0.276$.
There is a clear tendency in the autonomous model B to support x1v1-like
structures for larger perturbations and longer times than in model A, despite
the fact that initially, in  both cases, we have $\Delta$ x1v1 POs. A
difference we traced between the two models is the presence of a 5-periodic
orbit with tori close, around the initial conditions of x1v1. The projections of
these tori in the $(z,p_z)$ plane, for a quasiperiodic orbit with initial
conditions (0.160...,0.2,0,0), are depicted with red colour in
Fig.~\ref{mBsoses}b.
Although these 5 tori do not isolate the consequents of orbits in the
neighbourhood of x1v1, they delay their diffusion to larger volumes in phase
space, as  can be seen in Fig.~\ref{mBsoses}c from the accumulation of black
points around the position of the x1v1 PO.
Despite the fact that in 3D systems tori cannot isolate volumes of
phase pace, such situations definitely increase the importance of
$\Delta$ orbits, such as x1v1 in the specific case.

\item  \textbf{Orbits in the x1v2 neighbourhood:}
Radial perturbations of the U PO x1v2, lead in
general to weakly chaotic orbits. In a first approximation we can say
that the closer to the ICs of the
 PO we start integrating an orbit, the longer it
stays bar-supporting. However, we encounter also cases in which bar- and
non-barred suporting phases alternate during the 5~Gyr period.
The GALI$_2$ indicator of such orbits indicates in many cases
 a weakly chaotic character, while the orbits support of double
boxiness with X features embedded in their face-on views.

Vertical perturbations of x1v2 in the $p_z$-direction follow the behaviour we
encountered in all models. Namely, for $|p_{z_0}| < |p_{z_0}\rm{(x1v2)}|$ the
orbits are regular, with $\infty$-type envelopes in their side-on profiles,
while for $|p_{z_0}| > |p_{z_0}\rm{(x1v2)}|$ they become gradually chaotic.
\end{itemize}

\subsection{The non-autonomous case}
\label{mB_nonauto}

\subsubsection{FAST $GM_B$ INCREASE}
\label{mB_fmb}

Starting from the orbits existing in the autonomous case with
$GM_B$=0.13, for \ej=$-0.352$, we follow first their morphological
evolution when we have an increase of $GM_B$ from 0.13 to 0.52 within
5~Gyr. As for model A, also for model B, this means that the mass of
the bar quadruples.

\paragraph{Evolution of ICs of POs in the TD model:}
\label{mB_po}

\begin{enumerate}
\item The evolution of x1: The evolution of the PO x1 in model B is towards the
same shape as in the fast evolving model A (i.e.~from a x1-like to a 3:1-like
morphology, Fig.~\ref{mA_x1_fast}a), but faster. The 3:1-like shape is
continuously present for $t \gtrapprox 1.5$~Gyr, instead of 3.75~Gyr in model
A.

\item The evolution of x1v1: We find a remarkable persistence to a morphology
typical for quasiperiodic orbits around x1v1 in the autonomous model, especially
in the side-on projections, during the 5~Gyr period.  The face-on view, for $t
\lessapprox1.5$~Gyr, is slightly asymmetric towards a quasiperiodic rm21
morphology. With increasing time this effect weakens, although existing, and
eventually the orbit is elliptic-like for  $t \gtrapprox 3.75$~Gyr. Comparing
this evolution with our results in the low mass bar case of model A
(Fig.~\ref{mA_x1v1_fast}), we observe that the role of x1v1 is pronounced in
model B, in which initially the bar is 2.6 times more massive than in model A.
This is associated with the structure of the phase space in model B, which is
characterized by the presence of chains of stability islands around x1v1.
\item The evolution of x1v2:  The evolution of x1v2, in its side-on view, is
similar to that of x1v2 in model A, in the sense that it keeps the $\infty$-type
morphology of its envelope during the 5~Gyr period, becoming narrower with time.
The face-on view evolution is, like that of x1 in model B, from
x1-like-quasiperiodic to 3:1-like-quasiperiodic (Figs.~\ref{mA_x1_fast}a,b).
However, in model B, the 3:1 character is discernible already for $t \gtrapprox
1.5$~Gyr.
\end{enumerate}

\paragraph{Evolution of ICs of non-POs of the TI model in
the TD system:}

\begin{enumerate}
\item  Perturbations of x1: Radially perturbed x1 orbits evolve as x1 itself.
Namely, there is a transition from the x1-like to the 3:1-like morphology, that
occurs for times $t > 1.5$~Gyr. Nevertheless, whenever the applied perturbation
brings the initial conditions in the rm33 zone (the region of the three
stability island around the innermost island of x1 in Fig.~\ref{mBsoses}a), the
orbit evolves keeping its rm33-like shape  (as in the three right panels of
Fig.~\ref{mA_x1pert_rad}a) during the whole period of the 5~Gyr, reducing its
extent along the major axis of the bar with time.

The evolution of perturbed in the vertical direction x1 orbits, depends, in
general, on their ICs at the beginning of their evolution. A general description
of orbits with $0.03\leq z_0 \leq 0.15$ is that an initially ansae type face-on
morphology, is changing to one that can be described as a ``right parenthesis''.
The larger the $z_0$ perturbations, the later the time the transformation
occurs. The ansae shape is formed as the orbit librates between two
parentheses-like structures, a left and a right one, symmetric with respect to
major axis of the bar. Finally, the ``right parenthesis''-like one prevails in
the orbits we investigated. This can be seen in Fig.~\ref{mBf_x1pertv}a, where
we depict the face-on projections of the orbit with initial conditions
(0.16,0.15,0,0) and initial \ej=$-0.352$. The parentheses patterns strongly
resemble orbits belonging to the o1 family found in a strong bar model by
\citet[][their figure 17]{spa02b}. However, the orbit we present here is 3D. The
side-on projections, after an initial hybrid x1v1/x1v2-like morphology for
$t\leq 1.25$~Gyr, develop the standard $\infty$-shaped outline
(Fig.~\ref{mBf_x1pertv}b) with the tendency to become planar with increasing
time.  This latter morphology dominates in the evolution of all vertically
perturbed x1 orbits we discuss here, during the largest part of the 5~Gyr time
period. The evolution of the GALI$_2$ index (Fig.~\ref{mBf_x1pertv}c) is in
agreement with the morphological evolution of the orbit, revealing the
transition from an initially weakly chaotic behaviour to a more regular
evolution at later times. For even larger $z_0$ values, we find boxy face-on
projections combined with frown-smile, boxy, side-on views. For example, for
$z_0=0.2$, the double boxy morphology follows an initial (t$\leq$ 1.5~Gyr) phase
during which we find the imprint of the multiplicity 5 orbit to which belong
the five islands in Fig.~\ref{mBsoses}.
\begin{figure}
\begin{center}
\includegraphics[width=\columnwidth]{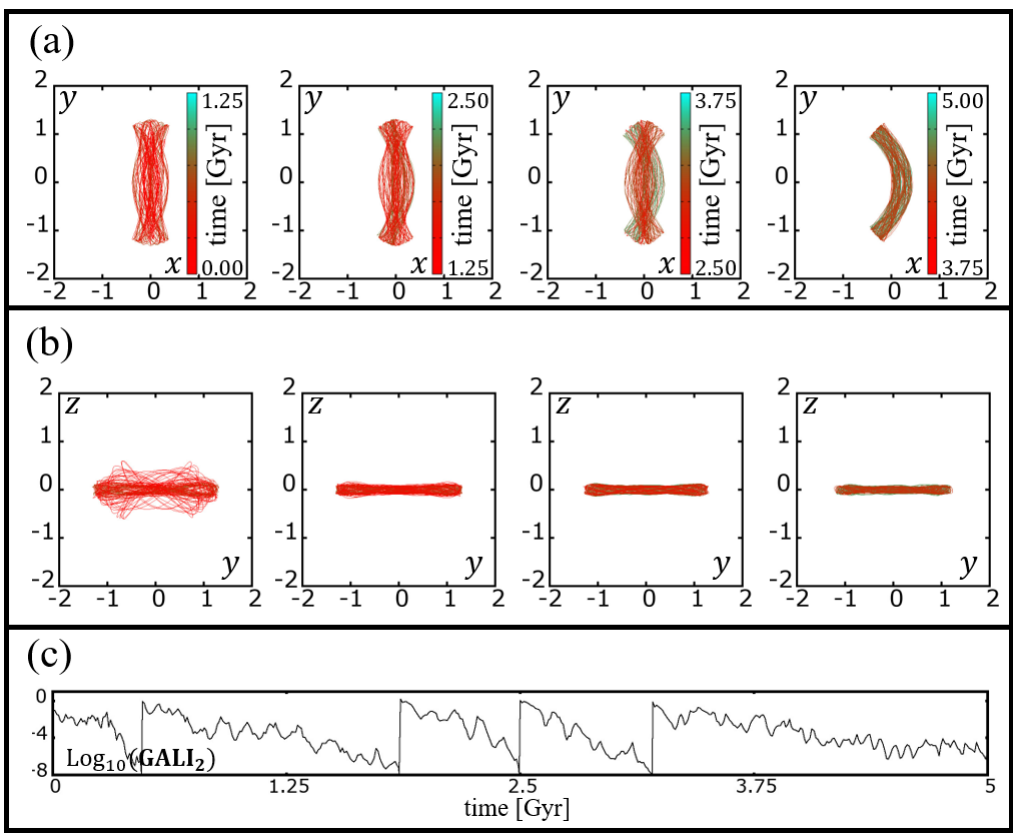}
\end{center}
\caption{Fast growing bar model B. Plots similar to
    Fig.~\ref{mA_x1v2_fast} but for a vertically perturbed x1 orbit,
    which shows a typical morphological evolution in its face-on view
    from an ansae type to a parenthesis-like morphology (a), while in
    its side-on view the $\infty$-shaped outline dominates (b).}
\label{mBf_x1pertv}
\end{figure}

\item  Perturbations of x1v1:
Both radial  and vertical perturbations of x1v1 lead to orbits that remain
closer to x1v1-like morphologies for larger perturbation ranges than in model A.
Radial perturbations in the range $0.13\leq x_0(\rm{x1v1})\approx 0.16 \leq
0.22$ keep their frown side-on shapes, while their face-on projections can be
briefly described as of distorted elliptical-like shapes. Deviations from the
$z_0 \approx 0.277$ IC keep their frown-like side-on morphologies during the
5~Gyr period, even if we reduce it to $z_0 = 0.18$, while for larger deviations
from $z_0$ in this direction, these morphologies, resembling those of
quasiperiodic orbits around x1v1, are transformed eventually to double boxy
ones.

\item Perturbations of x1v2:
The radially perturbed x1v2 orbits have a behaviour similar to the corresponding
orbits in model A. The main difference is that in the face-on views now dominate
the ansae-type shapes. In the side-on profiles we encounter again here the
$\infty$-type envelopes of the orbits during most of the time of the 5~Gyr long
integration interval. Their presence is associated with a more regular behaviour
of the orbits. Whenever this side-on morphology is distorted, the GALI$_2$ index
indicates a, weakly in general, chaotic character.

Vertically perturbed x1v2 orbits of model B, lead, as in model A,
either to the known orbits with the $\infty$-like
morphologies in their side-on views, as long as the perturbation
brings them on the tori of the 3D quasiperiodic orbits around x1, or
to the hybrid x1v1/x1v2 side-on views, when they are away from this
region. We note however, that in the latter case, frequently, weakly chaotic
and regular phases
alternate during the evolution of the orbits within 5~Gyr. This has as a result
a
rather constant support of the bar in the face-on view, combined
however with different degrees of support of a bar component by
different orbital shapes when viewed edge-on. We give an example in
Fig.~\ref{mBf_x1v2_ver} of an orbit with ICs
(0.141...,0,0,0.5). For $0\leq t \leq 1.25$~Gyr
  this orbit supports partly a bar. It has an irregular-boxy, face-on view
(Fig.~\ref{mBf_x1v2_ver}a) combined with a side-on profile harbouring an X
feature. It reaches heights about 0.8~kpc away of the equatorial plane.
In the time interval $1.25<t\leq2.5$~Gyr the orbit has a more regular shape with
a boxy face-on view and a clear smile-like, side-on one. Later, for
$2.5<t\leq3.75$~Gyr, the orbit behaves chaotically, returning to a morphology
similar to its initial at the beginning of the integration, while for the last
time interval ($3.75<t\leq5$~Gyr) we have a narrow side-on prfile, combined with
an asymmetric, bar-supporting, face-on projection. The GALI$_2$ index
clearly captures all these changes in the orbit's behaviour, as its several
reinitializations in the time intervals $0\leq t \leq 1.25$~Gyr and
$2.5<t\leq3.75$~Gyr indicate the orbit's chaotic nature, in contrast to its more
regular evolution for $1.25<t\leq2.5$~Gyr and $3.75<t\leq5$~Gyr. where the
reinitialization intervals became longer.
\begin{figure}
\begin{center}
\includegraphics[width=\columnwidth]{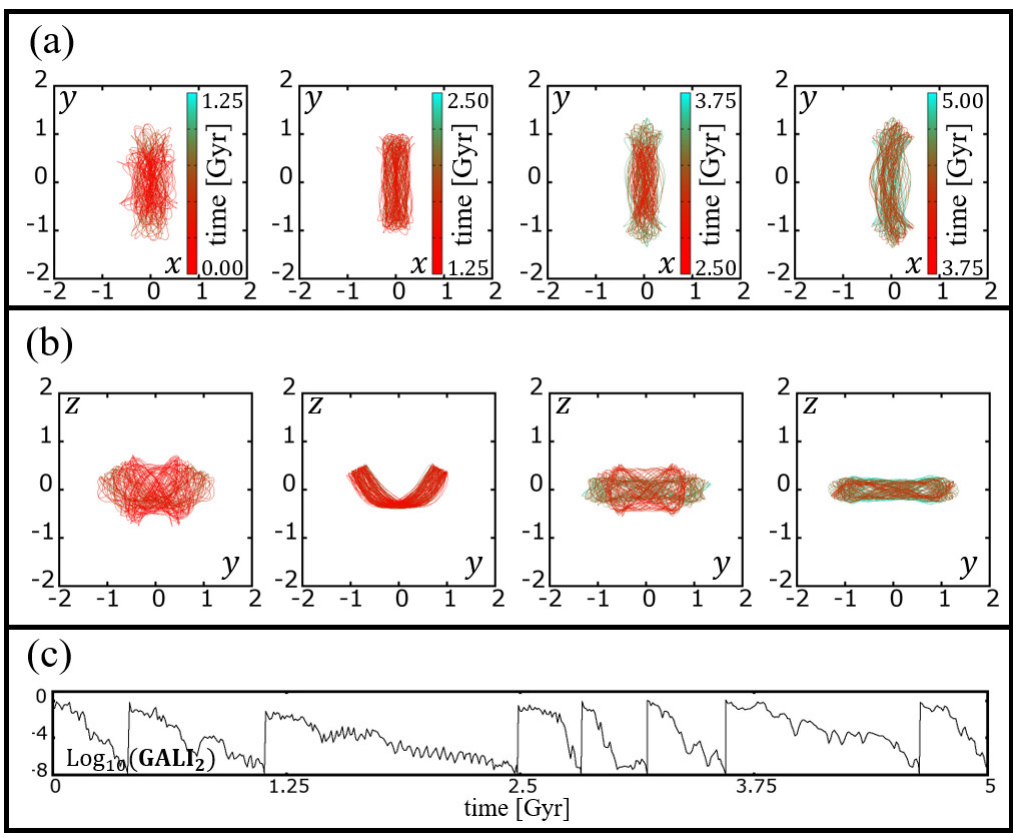}
\end{center}
\caption{Fast growing bar model B. Plots similar to
    Fig.~\ref{mA_x1v2_fast} but for a vertically perturbed x1v2 orbit. The orbit
is bar-supporting, but the supported bar morphologies, especially the side-on
profiles vary.}
\label{mBf_x1v2_ver}
\end{figure}
\end{enumerate}

\subsubsection{SLOW $GM_B$ INCREASE}
\label{mB_smb}

For the same model, B, we repeated our study considering that $GM_B$
increased within 5~Gyr from 0.13 to 0.156, i.e. by 20\%. Again at the
beginning of the integration the energy of the orbits we studied is
\ej=$-0.352$.

\paragraph{Evolution of ICs of POs in the TD model:}
\label{mB_spo}

\begin{enumerate}
\item The evolution of x1: In model B with $GM_B$ increasing slowly, as in the
corresponding model A, the shape of x1 remains invariant during the integration
period. There is no transition to 3:1-like shapes.

\item Also the shape of x1v1 remains quite invariant, despite being $\Delta$ at
the
starting point. This shape is characterized by a slightly triangular face-on
view, while the two other projections are similar to those of quasiperiodic
orbits around x1v1 in autonomous models (``thick'' frowns or smiles).

\item The evolution of the x1v2 PO, which is U in the starting autonomous
model, has also a regular character. The face-on projections resemble
quasiperiodic orbits around x1, while the side-on views reinforce the morphology
with the $\infty$-shaped outline. The same mechanism, which is being described
in Fig.~\ref{empty1}, is again in action. As a result, the evolution of an
unstable orbit in a TD model, with slowly increasing bar mass in this case,
leads to a regular behaviour.
\end{enumerate}

\paragraph{Evolution of ICs of non-POs of the TI model in the
TD system}
\begin{enumerate}

\item  Perturbations of x1: Radial perturbations of x1 evolve like quasiperiodic
orbits in the autonomous case, reflecting the shapes of the  POs, around which
they are trapped, i.e. of x1 and rm33.

The vertical perturbations of x1 evolve in a similar way like in the
case of model B with the fast increase of $GM_B$. We find a large
percentage of bar-supporting orbits with ansae-type face-on
projections, where the orbit evidently follows the two
parentheses-like shapes (cf.~Fig.~\ref{mBf_x1pertv}a), while the side-on
views are either of $\infty$-type, or hybrid x1v1/x1v2-like. For
perturbations beyond the $(z,p_z)$ region occupied by the x1 tori
(area between the two indicated with diamonds points in
Fig.~\ref{mBsoses}b), we find bar-supporting orbits with boxy face-on
projections, typically for $t\lessapprox2.5$~Gyr.
For longer integration times, the orbits are only partly
bar-supporting, or chaotic.

\item Perturbations of x1v1:
Perturbations of the x1v1 ICs in the
$x$-direction, in the range $0.13\leq x_0(\rm{x1v1})\approx 0.152\leq 0.22$,
retain a sharp frown-like shape in their side-on views, while in their
face-on projections we find patterns resembling quasiperiodic orbits
trapped around, or being sticky, to x1 or to its higher multiplicity
bifurcations \citep{path19}. Frown-like shapes persist in a considerable range
of perturbations in the $z$-direction.
\item Perturbations of x1v2:
We find bar-supporting orbits by perturbing x1v2 orbits both radially
and vertically. For perturbations in the $x$-direction, in the range
$0.12\leq x_0(\rm{x1v1})\approx 0.142 \leq 0.16$,  the elliptical shapes in
the face-on views, become boxy for $t
  \gtrapprox3.75$~Gyr. In the side-on views prevail the frown- or
smile-like patterns. This persistence of x1v1- or
 x1v1$^{\prime}$- shapes in the side-on projections
is also encountered in the vertical perturbations of x1v2 in this
model. Even orbits that start as weakly chaotic become later
bar-supporting with frown- or smile-like side-on views.  This
  is again related to the particular structure of phase space around
  the $\Delta$ x1v1 PO, which traps the orbits in a particular
  volume of it, as soon as the imposed perturbation brings its ICs in that
phase space region.

\end{enumerate}
\section{Model C: A model without Complex Unstable x1v1 orbits}
\label{mC}
The next model in which we have investigated the evolution of orbits as $GM_B$
increases, is model C. In this model the individual mass components are
$GM_{D}=0.878$, $GM_{S}=0.022$ and $GM_{B}=0.1$. The choice of these parameters
leads to a model, in which the important x1v1 family has no $\Delta$ parts.
The initial conditions of the main POs are for x1 (S)
$(x_0,z_0,p_{x_0},p_{z_0})\approx (0.154,0,0,0)$, for x1v1 (S) about
(0.15,0.144,0,0) and for x1v2 (U) about (0.149,0,0,0.076) at \ej=$-0.33$.

\subsection{The autonomous case}
\label{mC_autoc}

\begin{figure}
\begin{center}
\resizebox{85mm}{!}{\includegraphics[angle=270]{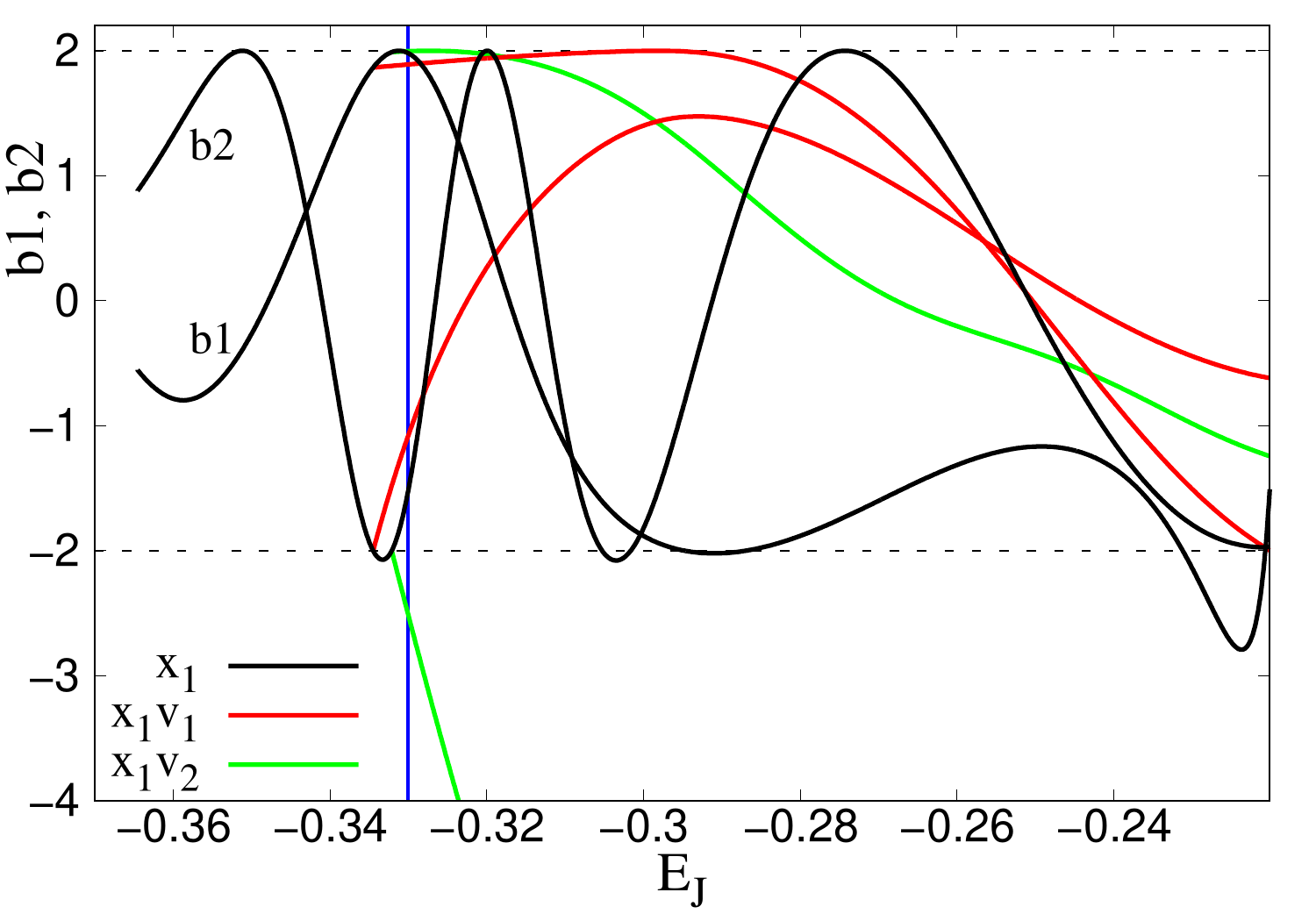}}
\end{center}
\caption{TI model C. Stability diagram similar to
    the one of Fig.~\ref{mA_stabdi}. We observe that x1v1 is always
    S. The vertical blue line indicates the energy value
    \ej=-0.33.}
\label{mCstab}
\end{figure}

\begin{figure}
\begin{center}
\includegraphics[width=\columnwidth]{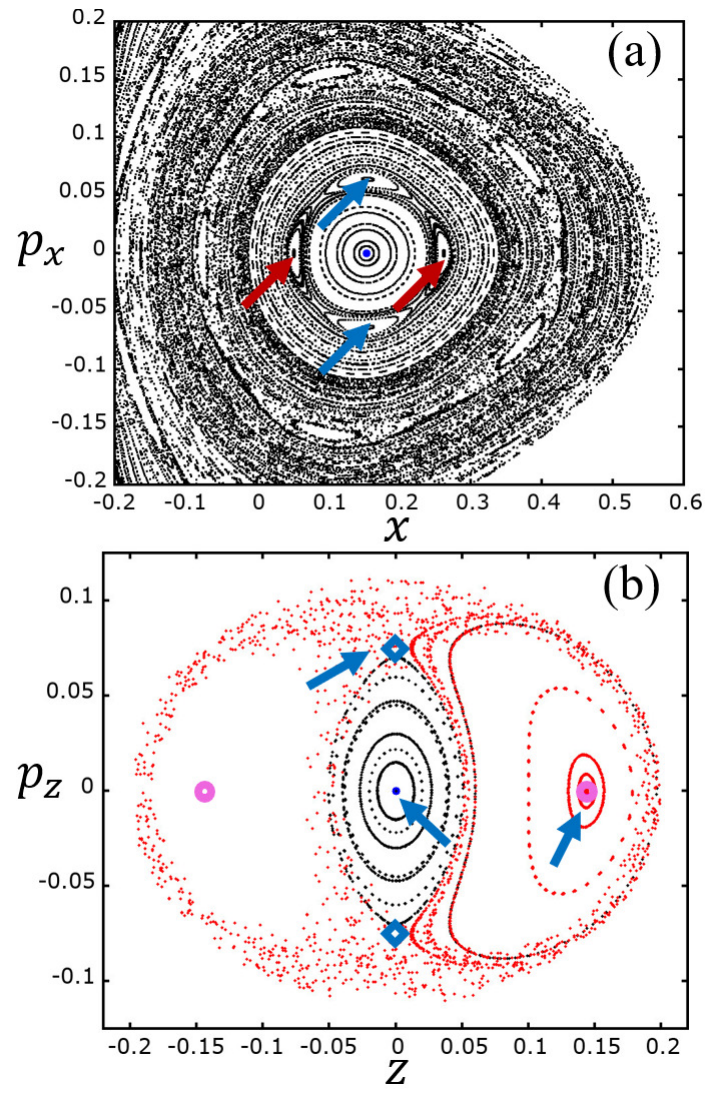}
\end{center}
\caption{TI model C for \ej=$-0.33$. (a) The
    $(x,p_x)$ surface of section around x1, located at the center of
    the innermost stability island (blue dot). Arrows
    point to the four islands of the multiplicity 2 POs, rm21 (red) and rm22
    (blue) (b) The $(z,p_z)$ projection of the Poincar\'{e} section
    around x1 (indicated by a blue dot and an arrow). Open circle
    symbols indicate the locations of x1v1 (right; also denoted by an
    arrow) and x1v1$^{\prime}$ (left), while the location of x1v2
    ($p_z>0$; also denoted by an arrow) and x1v2$^{\prime}$ ($p_z<0$)
    are indicated with ``diamond'' symbols. Black points correspond to
    vertical perturbations of x1, while red points belong  to several
    orbits with ICs in the neighbourhood of the x1v1 PO.}
\label{mCsoses}
\end{figure}

The property of the x1v1 family, for which we have chosen the model,
is demonstrated in the stability diagram given in
Fig.~\ref{mCstab}. We observe that the stability indices of this
family are always $-2<b1,2<2$, i.e.~its POs are
always S. We have chosen again an \ej to start with, at which all
three main families of POs coexist. This is
\ej=$-0.33$, for which x1 and x1v1 are S, while
x1v2 is U. Guidelines for understanding the
phase space structure at \ej=$-0.33$, can be taken again by plotting
the $(x,p_x)$ surface of section for the planar orbits
(Fig.~\ref{mCsoses}a) and the $(z,p_z)$ projection of the 4D
Poincar\'{e} cross section (Fig.~\ref{mCsoses}b).

\begin{itemize}
 \item \textbf{Orbits in the x1 neighbourhood:}
As we can see in Fig.~\ref{mCsoses}a, the $(x,p_x)$ surface of section in the x1
region is characterized by islands of stability belonging to higher multiplicity
bifurcations of x1. There is little chaos in the x1 neighbourhood. Consequently,
by perturbing x1 radially, we displace the ICs of the orbit we want to study  on
some invariant curve around x1 or on an invariant curve belonging to a stability
island of another family (e.g. on the four islands of rm21/rm22, denoted
respectively by red/blue arrows). Practically, we always have a regular
bar-supporting orbit, either by considering a single quasiperiodic orbit, or
two, symmetric orbits, as e.g. in the case of rm21 and rm22.

Also by applying vertical perturbations to x1, we reach tori of quasiperiodic
orbits. By perturbing x1 in the $p_z$-direction, we find, as $|p_z|$ increases,
for $|p_z| < |p_{z_0}(\rm{x1v2})|$ again the orbits with the expected
$\infty$-shaped side-on profiles of the invariant-like curves around x1 in the
central region of Fig.~\ref{mCsoses}b. For larger $|p_z|$ perturbations, hybrid
x1v1/x1v2-like side-on projections substitute the ``$\infty$'' shape profiles.

\item \textbf{Orbits in the x1v1 neighbourhood:}
In the absence of ($\Delta$) regions, radial and vertical perturbations of x1v1
orbits lead in the first place to quasiperiodic orbits around them, having the
expected morphologies of such orbits encountered in autonomous models. Namely,
we have frown-like quasiperiodic orbits around x1v1 and smile-like quasiperiodic
orbits around x1v1$^{\prime}$. This is expected, since in model C we do not have
any kind of a chaotic zone around x1v1 and x1v1$^{\prime}$, but invariant tori
surrounding them in the 4D space of section, as those described in
\citep{kp11}.

\item  \textbf{Orbits in the x1v2 neighbourhood:} Radial and vertical
perturbations of x1v2 lead to  morphologies that can be inferred by considering
its initial conditions on the projections of the surfaces of section depicted in
Fig.~\ref{mCsoses}. The difference with respect to the previous studied cases,
is that in the region surrounding x1v1 and x1v1$^{\prime}$ are occupied by tori
on which perturbed x1v2 orbits may become sticky.
\end{itemize}

\subsection{The non-autonomous case}
\label{mC_nonauto}

\subsubsection{FAST $GM_B$ INCREASE}
\label{mC_fmb}

Starting from the orbits existing in the autonomous case for \ej=$-0.33$, we
follow first their morphological evolution when we have an increase of $GM_B$
from 0.1 to 0.4 within 5~Gyr. Like in the two previous models, this means that
the mass of the bar quadruples.

\paragraph{Evolution of ICs of POs in the TD model:}
\label{mC_po}

\begin{enumerate}
 \item The evolution of x1: The evolved x1 orbit
   remains practically invariant as in the autonomous case.
 \item The evolution of x1v1: The x1v1 orbit keeps
   a x1v1-like morphology of a quasiperiodic orbit close to a periodic
   one. However, there is a clear tendency of the extent of the
   x1-like ellipse in the face on views to shrink with time. Namely,
   its projection on the major axis of the bar, the y-axis, becomes
   shorter.
 \item  The evolution of x1v2: The x1v2 orbit evolves similar to x1 orbits,
perturbed in the $p_z$ direction. It has a quasiperiodic x1-like face-on and the
known ``$\infty$'' side-on, overall shape. As in many previous cases we
presented up to now, also in model C, the side-on profile becomes more narrow
with time.
\end{enumerate}

\paragraph{Evolution of ICs of non-POs of the TI model in the TD
system:}

\begin{enumerate}
\item  Perturbations of x1: Radial perturbations of x1 have always morphologies
similar to quasiperiodic orbits around the PO of the autonomous case. They are
elliptical-like, if located close to the x1 and form boxy structures, if their
ICs are further away (alone or in symmetric pairs).

Orbits that are vertical perturbations of x1 have in general morphologies
resembling in their side-on views either quasiperiodic orbits around
x1v1/x1v1$^{\prime}$ (``thick'', frown- or smile-like morphologies), or 3D,
quasiperiodic orbits around x1 with overall $\infty$-like shapes. Their
morphology is determined by the location of their ICs on the Poincar\'{e}
sections of the shadow models.

\item Perturbations of x1v1: Radially and vertically perturbed ICs of the x1v1
PO lead to orbits with x1v1-like shapes. Varying $x_0$ in the $0.13\leq x_0\leq
0.22$ range we found orbits with edge-on quasiperiodic-x1v1-like morphologies,
while in their face-on views we find in most cases the elliptical-like
quasiperiodic x1 patterns. During their evolution we also encounter
quasiperiodic rm21/rm22 morphologies in  some time  intervals.

Vertically perturbed x1v1 orbits have similar behaviours with the vertically
perturbed x1 orbits. They keep qualitatively their initial quasiperiodic x1v1
morphology for 5~Gyr, shrinking however along the major axis of the bar with
time. Transition to double boxy morphologies during the integration time is
observed when we start with ICs in the borders of the regions occupied by x1 and
x1v1 tori (Fig.~\ref{mCsoses}b.)

\item Perturbations of x1v2: Starting integrating orbits in the neighbourhood of
x1v2 we find in general regular orbits. In order to find the tendencies for the
morphological evolution of perturbed x1v2 orbits in the TD model
Fig.~\ref{mCsoses}b is proven again a useful tool. Starting from x1v2 and moving
in the $p_z$-direction towards x1, we find the orbits with elliptical-like
face-on and $\infty$-like side-on profiles.

Regular orbits are encountered in the TD model C even away from the region with
the main families of POs. Such orbits are characterized by boxy edge-on profiles
with  a possible profile evolution in the face-on views, where we have
transitions from elliptical-like to boxy patterns. A typical example of  this
behaviour, for an orbit with ICs (0.148...,0,0,0.2) is given in
Fig.~\ref{mCf_x1v2pertv}. The face-on transition takes place for $t>3.75$
(Fig.~\ref{mCf_x1v2pertv}a). The time evolution of the GALI$_2$ index
(Fig.~\ref{mCf_x1v2pertv}c) clearly indicates the regular behavior of the orbit
as  GALI$_2$ remains always larger than the threshold value $10^{-8}$. There is
a characteristic difference with respect to boxy profiles we found in the others
models, namely that they are not densely filled, probably because of trapping
around quasiperiodic orbits of high multiplicity.
\begin{figure}
\begin{center}
\includegraphics[width=\columnwidth]{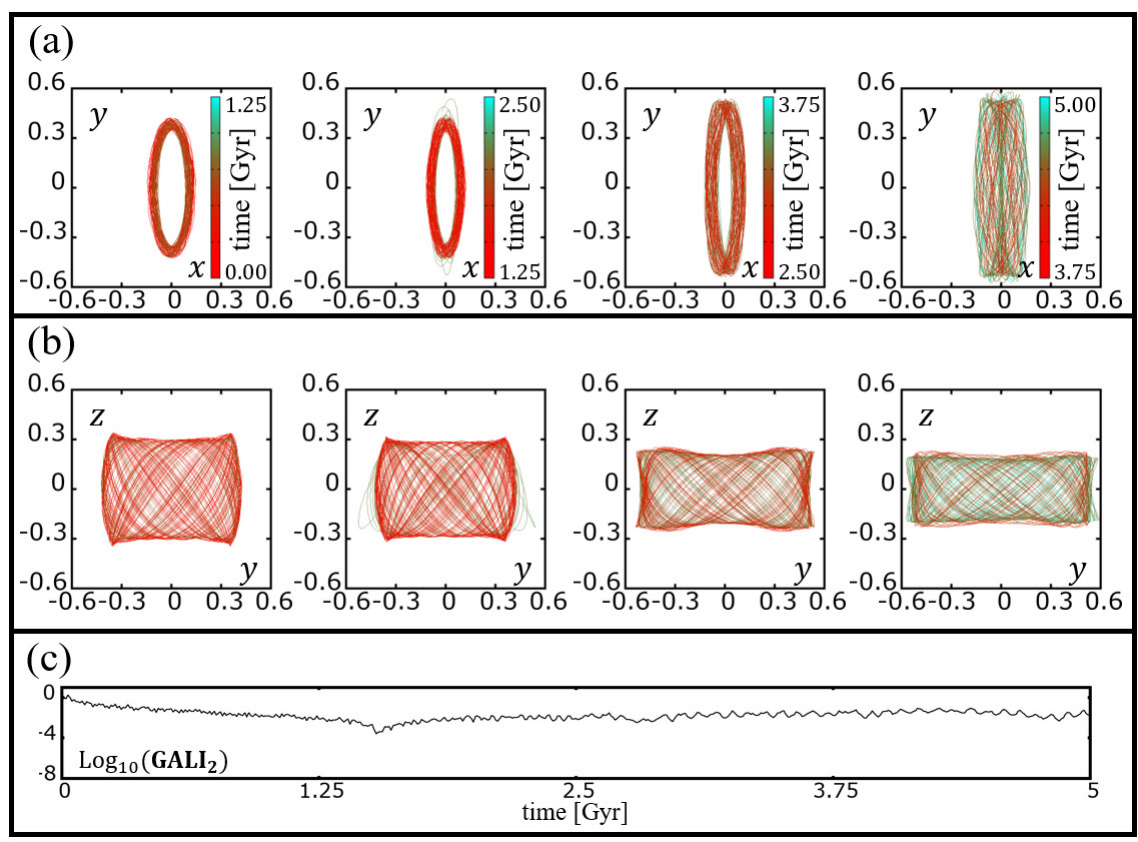}
\end{center}
\caption{Fast growing bar model C. Plots similar to
  Fig.~\ref{mA_x1v2_fast} but for a vertically perturbed x1v2 orbit, with ICs
displaced beyond the region of the neighbourhod of the three main families of
the ILR region. The orbit has a regular character with boxy side-on profiles
(see text).}
\label{mCf_x1v2pertv}
\end{figure}
\end{enumerate}

\subsubsection{SLOW $GM_B$ INCREASE}
\label{mC_smb}

For model C, we repeated our previous study considering that $GM_B$ increases
from 0.1 to 0.12, i.e. by 20\%, within 5~Gyr. Also for the study of the slowly
increasing $GM_B$ the considered orbits have initially \ej=$-0.33$.

\paragraph{Evolution of ICs of POs in the TD model:}
\label{mC_spo}

The x1 and x1v1 orbits remain invariant during the 5~Gyr integration time. They
retain their known morphologies as in the TI case.  The orbit starting with the
ICs of x1v2, has continuously a shape similar of an orbit on the invariant tori
around x1 in the $(z,p_z)$ projections of the surface of section. This is a
morphology similar to the one in Fig.~\ref{tigalis3a}a, with elliptical, x1-like
face-on projections, which remain as such during the whole integration period.

\paragraph{Evolution of ICs of non-POs of the TI model in the TD
system:}
\label{mC_snpo}

\begin{enumerate}
\item  Perturbations of x1:  Perturbations of the x1 PO ($x_0(\rm{x1})\approx
0.154$) in
the x-direction, lead to elliptical-like quasiperiodic orbits for
$\Delta x_0 \lessapprox 0.1$ and then become boxy. However, whenever the
starting point is located in the vicinity of a regular PO of higher
multiplicity, e.g.~on one of the rm21/rm22 islands, a shape similar to that PO
is preserved for the whole integrating period.

Vertically perturbed x1 orbits remain in general regular in this
model, with shapes similar to those of the vertically perturbed x1
orbits in the fast growing bar case of model C. Their evolution can be
inferred by the morphology of the orbit at the starting point of the
autonomous model, which remains practically invariant.

\item  Perturbations of x1v1: All perturbations of x1v1 in the slowly growing
bar model C, lead to regular orbits with quasiperiodic-like morphologies
resembling quasiperiodic orbits around x1v1 in the autonomous case. Whenever a
morphological evolution is observed, the traces of the orbit in the Poincar\'{e}
sections of the corresponding autonomous models of the shadow evolution are
found beyond the region occupied by the tori around x1v1 (or x1v1$^{\prime}$),
as they appear in $(z,p_z)$ projections similar to Fig.~\ref{mCsoses}b. Such
transitions are happening less frequently in the slowly increasing $GM_B$ model,
because the deviations of the evolved orbits from the orbit of the TI model with
the same ICs are small and the phase-space of the model is dominated by order.

\item  Perturbations of x1v2: Radially perturbed x1v2 orbits, as long as their
ICs are located beyond the central, lemon-shaped area, with the x1v2 and
x1v2$^{\prime}$ POs at its peaks (Fig.~\ref{mCsoses}b), have side-on views
either with boxy morphologies, or with quasiperiodic-frown/smile-like shapes.
The latter happens when the imposed perturbations bring them in the zone of
influence of x1v1 (or x1v1$^{\prime}$). There are cases, for which we have a
drift from an initial boxy morphology to a quasiperiodic-x1v1-like one, during
the last phases of an orbital evolution. This reflects the gradual shift of the
trace of the orbit in the shadow models, as $GM_B$ increases, to regions
occupied by x1v1 tori. We note that the orbits, which evolve morphologically,
show in general a delay in their evolution in the slowly growing bar case,
compared with the fast growing one.

Examining vertically perturbed x1v2 orbits ($p_z$(x1v2)$\approx 0.076$) in the
$p_z$-direction, with $0.076 \leq p_z \leq  0.2$, we always find boxy side-on
profiles, similar to those of the perturbed x1v2 orbits of the fast growing bar
case (Fig.~\ref{mCf_x1v2pertv}b). During several time windows, these boxy shapes
point clearly to side-on views of higher multiplicity periodic orbits as those
presented in \citet{path19}.
\end{enumerate}

\section{Summary and Conclusions}
\label{sec:concl}
We have studied the evolution of specific orbits in Ferrers bars, in which the
mass of the bar increases with time. We have examined two rates of bar growth
within 5~Gyr. A fast one, in which the bar mass quadruples and a slow one, in
which at the end of the integration the mass of the bar has increased just by
20\%.

The orbits we investigated are those, which in TI models contribute mostly to
the orbital content of the thick part of the bar, i.e. of the b/p bulge. This is
the reason we concentrated ourselves in the study of the main orbital families,
which provide the building blocks for thick bars. These are the
families x1, x1v1 and x1v2, which start coexisting at the ILR region of the
models. By using orbits associated with them, one can build profiles
reproducing the observed edge-on morphology of boxy bulges, which are
considered to be the part of the bar that extends well above the equatorial
plane of the galaxy \citep[see e.g.][]{ath05, betal06, ath16, pxaa21}.

Our goal is to find out the degree at which orbits starting with the same
initial conditions as those in the autonomous case, continue to support a bar as
the mass of the bar component of the potential increases. For this purpose it is
important to know  the regular or chaotic character of the orbits, although this
fact does not secure by itself that they are bar-supporting or not. For
determining the regular or chaotic nature of orbits during their time evolution
we have used the GALI$_2$ index. In our study, we have also used  (a) stability
diagrams in the initial, autonomous models, which describe the evolution of the
linear stability of the three main families of periodic orbits, (b) the
$(x,p_x)$ Poincar\'{e} sections, as well as appropriate projections in the
$(z,p_z)$ plane of the 4D cross sections of the autonomous case at the starting
energy and (c) the shadow evolution of each model. This is a series of
successive autonomous, ``frozen'', models with the $GM_B$ values taken by the
orbit we study during its integration.

The main conclusions of our study are:
\begin{enumerate}
\item The bar-supporting orbital morphologies found in TI models to support a
b/p side-on profile, are encountered also in models, in which the mass of the
bar increases with time. We did not find any orbital shapes not existing in
autonomous models, which reinforce the peanut-shaped part of the bar.
\item The morphological evolution of the orbits observed in the TD models is
gradual, even in the cases in which the mass of the bar quadruples within the
integration time (5~Gyr). Contrarily to autonomous models, where we find
variations in orbital morphologies only in the neighbourhood of unstable POs,
the morphology of orbits in TD models may also vary in cases where an orbit
remains practically regular during the integration time. In the majority of the
studied cases, when a bar-supporting orbit ceases providing bar-supporting
shapes, it becomes and remains chaotic for the rest of the time. Nevertheless,
we also found cases, where bar-supporting and no-bar-supporting phases
alternate
(e.g. Fig.~\ref{mBf_x1v2_ver}).
\item There is a good agreement between the morphology of an orbit in a time
window and the location of its traces on the Poincar\'{e} cross section at the
corresponding \ej in the autonomous model with the $GM_B$ value of the orbit at
the time we are interested in. Thus, in a first approximation, we can understand
the morphological evolution of an orbit in the TD potential as the series of
morphological transformations expected by the displacements of its traces on the
Poincar\'{e} sections during the shadow evolution. This applies especially to
models with less chaos, as e.g. in the case of model C.
\item The role of the planar x1 orbits in the building of the 3D part of the
bars has been emphasized in \citet{pk14a, pk14b} and in \citet{cppsm17}. The
present study shows that this role can be even more important in TD models. The
mechanism that leads to the reinforcement of the thick part of the bar, is
based on the 3D morphology of the
orbits on the tori around x1 in the 4D space of section and is in favor of
``CX''  \citep{betal06} side-on profiles. The orbits on these tori, with
$x_0=x_0(\rm{x1})$, $|p_{z_0}|< |p_{z_0}(\rm{x1v2})|$ and $p_{x_0}=z=0$, are
regular, with an $\infty$-type outline. However, in the autonomous models, if we
have $|p_{z_0}|\geqq |p_{z_0}(\rm{x1v2})|$ we find initially sticky chaotic, and
for larger $|p_{z_0}|$ chaotic, orbits. Such orbits, either offer a weaker
support to the bar with hybrid x1v1/x1v2 side-on profiles, or no support at all.
In the growing bar models we studied, with increasing time, an initial
$|p_{z_0}|$ value with $|p_{z_0}| \gtrapprox |p_{z_0}(\rm{x1v2})|$, drifts
towards smaller values and the orbit has a regular behaviour, evolving as
belonging to quasiperiodic orbits in an autonomous model. A typical example is
given in Fig.~\ref{empty1}.
\item  The regular or chaotic character of an orbit depends strongly on the
structure of the phase space around it and in the way this phase space
environment
evolves during the time of integration. Thus, an initially quasiperiodic orbit,
in a stability island  surrounded by a chaotic sea, may evolve mainly as
chaotic in the TD model. In parallel, an initially chaotic orbit, close to an
unstable PO, in the TI model, may behave as regular, if there are stability
islands of other orbits in its neighbourhood. The regular behaviour of orbits in
the neighbourhood of simple unstable x1v2 orbits in all models with increasing
in time $GM_B$ and partly the regular behaviour of orbits close to the complex
unstable x1v1 orbit in model B, are due to the structure of the phase space
around the POs.
\item  Comparing the various studied models among themselves, we draw
the following conclusions:

(a) By evolving the same ICs, we find less
morphological evolution in slowly than in fast growing $GM_B$  models.

(b) We
find more bar-supporting orbits in model B, in which we start with a more
massive bar than in model A, in which the mass of the bar is at the beginning
less massive. This is determined by the structure of phase space, especially
around x1v1 (Fig.~\ref{mBsoses}).

(c) In model B with the massive bar, we found
more planar orbits in the ILR region that support boxy shapes, than in the two
other models. Vertical perturbations of such orbits retain boxy projections on
the equatorial plane for a considerable range of the perturbations.

(d) The ILR
region of model C, with a marginal spheroidal component, is characterized by the
dominance of large stability areas in phase space. This favours the persistence
of structures like b/p bulges supported by regular orbits, as the mass of the
bar increases with time.

\end{enumerate}
In our time dependent models with the growing in time mass of the bar,
time-dependency has contributed to the regularization of orbital motion, at
least in the important for our study case of orbits in the vicinity of the
initially unstable POs of the x1v2 family. This introduces a new way of
supporting observed bar morphologies and particularly bars with ``CX''-type
side-on profiles. Future work should investigate whether this mechanism is in
action in $N$-body models, especially during the growth phase of the bar.

\vspace{0.5cm}
\noindent \textit{Acknowledgements}
We thank Prof.~G.~Contopoulos for fruitful discussions and very useful
comments.  Ch.S.~acknowledges support by the
  Research Committee (URC) of the University of Cape Town, and thanks
  the Research Center for Astronomy and Applied Mathematics of the Academy
of Athens for its hospitality during his visits when
  parts of this work were carried out. This work has been been partially
supported by the Research Committee of the Academy of Athens through
the project 200/895.

\vspace{0.5cm}
\textbf{Data availability}
The data underlying this article will be shared on reasonable request to the
corresponding author.

\label{lastpage}


\begin{thebibliography}{}

\bibitem[\protect\citeauthoryear{Athanassoula et al.}{1983}]{a83}
  Athanassoula E., Bienayme O., Martinet L., Pfenniger D. 1983, A\&A,
  127, 349

\bibitem[\protect\citeauthoryear{Athanassoula}{2003}]{ath03} Athanassoula E.,
2003, MNRAS 341, 1179

\bibitem[\protect\citeauthoryear{Athanassoula}{2005}]{ath05} Athanassoula E.,
2005, MNRAS 358, 1477

\bibitem[\protect\citeauthoryear{Athanassoula}{2013}]{ath13} Athanassoula E.,
2013, in ``Secular Evolution of Galaxies'', Falcon-Barroso J., and Knapen J.H.
(eds.), Cambridge, UK: Cambridge University Press, 2013, p.305

\bibitem[\protect\citeauthoryear{Athanassoula}{2016}]{ath16} Athanassoula E.,
2016, in "Galactic Bulges", E. Laurikainen, R. Peletier, D. Gadotti, (eds.),
Springer International Publishing Switzerland, pp. 391-412

\bibitem[\protect\citeauthoryear{Benettin et al.}{1980a}]{benetal80a}
   Benettin G., Galgani L., Giorgilli A., Strelcyn
    J.-M., 1980a, Mecc, 15, 9

\bibitem[\protect\citeauthoryear{Benettin et al.}{1980b}]{benetal80b}
   Benettin G., Galgani L., Giorgilli A., Strelcyn
    J.-M., 1980b, Mecc, 15, 21

\bibitem[\protect\citeauthoryear{Broucke}{1969}]{b69} Broucke R.,
  1969, NASA Techn. Rep., 32, 1360

\bibitem[\protect\citeauthoryear{Bureau et al.}{2006}]{betal06} Bureau
  M., Aronica G., Athanassoula E., Dettmar R.-J., Bosma A., Freeman
  K. C., 2006, MNRAS, 370, 753

\bibitem[\protect\citeauthoryear{Chatzopoulos et
al.}{2011}]{cpb11} Chatzopoulos S., Patsis P. A., Boily C.M., 2011,  MNRAS,
416, 479

\bibitem[\protect\citeauthoryear{Chaves-Velasquez et
    al.}{2017}]{cppsm17} Chaves-Velasquez L., Patsis P. A., Puerari
  I., Skokos Ch., Manos T. 2017, ApJ, 850, 145

\bibitem[\protect\citeauthoryear{Combes et al.}{1990}]{cbfp90} Combes
  F, Debbasch F., Friedli, D, Pfenniger D., 1990, A\&A, 233, 82

\bibitem[\protect\citeauthoryear{Contopoulos}{2004}]{gcobook}
  Contopoulos G., 2004, ``Order and Chaos in Dynamical Astronomy'',
  Springer-Verlag, Berlin, Heidelberg, New York

\bibitem[\protect\citeauthoryear{Contopoulos \& Grosb{\o}l}{1989}]{cg89}
Contopoulos G., Grosb{\o}l P., 1989, A\&ARv, 1, 261

\bibitem[\protect\citeauthoryear{Contopoulos \& Harsoula}{2008}]{ch08}
Contopoulos G., Harsoula M., 2008, Int. J. Bifurc. Ch. 18, 2929

\bibitem[\protect\citeauthoryear{Contopoulos \& Magnenat}{1985}]{cm85}
Contopoulos G., Magnenat P., 1985, Celest. Mech. 37, 387

\bibitem[\protect\citeauthoryear{Contopoulos \&
    Papayannopoulos}{1980}]{cp80} Contopoulos G., Papayannopoulos T.,
  1980, A\&A, 92, 33

\bibitem[\protect\citeauthoryear{Ferrers}{1877}]{f887}
Ferrers, N. M. 1877, Quart.J.Pur.Appl.Math., 14, 1

\bibitem[\protect\citeauthoryear{Hadjidemetriou}{1975}]{hdj75}
  Hadjidemetriou J., 1975, Celest. Mech., 12, 255

\bibitem[\protect\citeauthoryear{Harsoula \& Kalapotharakos}{2009}]{hk09}
Harsoula M., Kalapotharakos C., 2009, MNRAS, 394, 1605

\bibitem[\protect\citeauthoryear{Katsanikas \& Patsis}{2011}]{kp11}
  Katsanikas M., Patsis P.A., 2011,  Int. J. Bif. Ch., 21-02, 467

\bibitem[\protect\citeauthoryear{Katsanikas et al.}{2011}]{kpc11}
  Katsanikas M., Patsis P.A., Contopoulos G., 2011, Int. J. Bif. Ch.,
  21, 2321

\bibitem[\protect\citeauthoryear{Katsanikas et al.}{2013}]{kpc13}
  Katsanikas M., Patsis P.A., Contopoulos G., 2013, Int. J. Bif. Ch.,
  23-02, 1330005

\bibitem[\protect\citeauthoryear{Lange et al.}{2014}]{LROBK14}
  Lange S., Richter M., Onken F.,  B\"{a}cker A.,
    Ketzmerick R., 2014, Chaos, 24, 024409

\bibitem[\protect\citeauthoryear{Machado \& Athanassoula}{2010}]{ma10}
 Machado R.E.G., Athanassoula E., 2010, MNRAS, 406, 2386

\bibitem[\protect\citeauthoryear{Machado \& Manos}{2016}]{mm16}
 Machado R.E.G., Manos T., 2016, MNRAS, 458, 3578

\bibitem[\protect\citeauthoryear{Manos \& Machado}{2014}]{mm14}
Manos T., Machado R.E.G., 2014, MNRAS, 438, 2201

\bibitem[\protect\citeauthoryear{Manos et al.}{2012}]{MSA12}
   Manos T., Skokos Ch., Antonopoulos Ch., 2012,
    Int. J. Bif. Ch., 22, 1250218

\bibitem[\protect\citeauthoryear{Manos et al.}{2013}]{mbs13} Manos T.,
Bountis, T., Skokos, Ch., 2013, Journal of Physics A, 46, 25-4017

\bibitem[\protect\citeauthoryear{Miyamoto \& Nagai}{1975}]{mn75}
  Miyamoto M., Nagai R., 1975, PASJ, 27, 533

\bibitem[\protect\citeauthoryear{Moges et al.}{2020}]{MMS20}
   Moges H., Manos T., Skokos Ch., 2020,
    Nonlin. Phenom. Complex Syst., 23, 153

\bibitem[\protect\citeauthoryear{Onken  et al.}{2016}]{OLKR16}
   Onken F., Lange S., Ketzmerick R., B\"{a}cker A.,
    2016, Chaos, 26, 063124

\bibitem[\protect\citeauthoryear{Patsis}{2005}]{p05} Patsis P.A.,
2005, MNRAS, 358, 305

\bibitem[\protect\citeauthoryear{Patsis et al.}{1997}]{paq97} Patsis P. A.,
Athanassoula E., Quillen A. C., 1997, ApJ, 483, 731

\bibitem[\protect\citeauthoryear{Patsis \& Athanassoula}{2019}]{path19} Patsis
P.A., Athanassoula E., 2019, MNRAS, 490, 2740%

\bibitem[\protect\citeauthoryear{Patsis \& Harsoula}{2018}]{ph18}
  Patsis P.A., Harsoula M., 2018, A\&A,  612, 114

\bibitem[\protect\citeauthoryear{Patsis \& Katsanikas}{2014a}]{pk14a}
  Patsis P.A., Katsanikas M., 2014, MNRAS, 445, 3525

\bibitem[\protect\citeauthoryear{Patsis \& Katsanikas}{2014b}]{pk14b}
  Patsis P.A., Katsanikas M., 2014, MNRAS, 445, 3546

\bibitem[\protect\citeauthoryear{Patsis et al.}{2002}]{psa02}
   Patsis P.A., Skokos Ch., Athanassoula E., 2002, MNRAS, 337, 578

\bibitem[\protect\citeauthoryear{Patsis et al.}{2021}]{pxaa21}
   Patsis P.A.,  Xilouris E.M., Alikakos J., Athanassoula E., 2021, A\&A, 647,
20

\bibitem[\protect\citeauthoryear{Patsis \& Zachilas}{1994}]{pz94}
  Patsis P.A., Zachilas L., 1994, Int. J. Bif. Ch., 4, 1399

\bibitem[\protect\citeauthoryear{Pfenniger}{1984}]{pf84} Pfenniger D.,
  1984, A\&A, 134, 373

\bibitem[\protect\citeauthoryear{Plummer}{1911}]{pl11} Plummer H.C., 1911,
MNRAS, 71, 460

\bibitem[\protect\citeauthoryear{Poincar\'{e}}{1899}]{poin99}
  Poincar\'{e} H., 1899, Les Methodes Nouvelles de la Mechanique
  Celeste, Vol. III, Gauthier-Villars, Paris

\bibitem[\protect\citeauthoryear{Richter et
    al.}{2014}]{RLBK14} Richter M., Lange S.,
  B\"{a}cker A., Ketzmerick R., 2014, Phys.~Rev.~E, 89, 022902

\bibitem[\protect\citeauthoryear{Skokos}{2001}]{S01}
   Skokos Ch., 2001, Physica D, 159, 155

\bibitem[\protect\citeauthoryear{Skokos}{2010}]{S10}
   Skokos Ch., 2010, Lect.~Notes Phys., 790, 63

\bibitem[\protect\citeauthoryear{Skokos \& Manos}{2016}]{sm16} Skokos Ch.,
Manos T., 2016, Lect.~Notes Phys., 915, 129

\bibitem[\protect\citeauthoryear{Skokos et al.}{2007}]{SBA07}
   Skokos Ch.,  Bountis T.C., Antonopoulos Ch.,
    2007, Physica D, 231, 30

\bibitem[\protect\citeauthoryear{Skokos et al.}{2008}]{SBA08}
   Skokos Ch.,  Bountis T.C., Antonopoulos Ch.,
    2008, Eur. Phys. J. Sp. Top., 165, 5

\bibitem[\protect\citeauthoryear{Skokos et al.}{2002a}]{spa02a} Skokos
  Ch., Patsis P.A., Athanassoula E., 2002a, MNRAS, 333, 847

\bibitem[\protect\citeauthoryear{Skokos et al.}{2002b}]{spa02b} Skokos
  Ch., Patsis P.A., Athanassoula E., 2002b, MNRAS, 333, 861

\bibitem[\protect\citeauthoryear{Tsigaridi \& Patsis}{2015}]{tp15}
  Tsigaridi L., Patsis P.A., 2015, MNRAS,  448, 3081

\end{thebibliography}
\end{document}